\documentclass[prd,preprint,tightenlines,showpacs,preprintnumbers,nofootinbib,eqsecnum,superscriptaddress]{revtex4-1}

 \usepackage[dvips,final]{graphicx}
  \usepackage{amssymb}
   \usepackage{amsmath}
    \usepackage{amsfonts}
     \usepackage{epsfig}
      \usepackage{bm}

\usepackage{mathpazo}

\usepackage[section]{placeins}

\usepackage{sidecap}
\usepackage{multirow}
\usepackage{ctable}
\usepackage{booktabs}
\usepackage{array}
\usepackage{tabularx}
\usepackage{xcolor}
\usepackage{pstricks}

\renewcommand{\thetable}{\arabic{table}}
\begin{document}


\title{Associated production of $\chi_c$ pairs with a gluon 
	in the collinear-factorization approach}

\author{Izabela Babiarz}
\email{izabela.babiarz@ifj.edu.pl.pl}
\affiliation{Institute of Nuclear Physics, Polish Academy of Sciences, 
ul. Radzikowskiego 152, PL-31-342 Krak{\'o}w, Poland}

\author{Wolfgang Sch\"afer}
\email{Wolfgang.Schafer@ifj.edu.pl} 
\affiliation{Institute of Nuclear
Physics, Polish Academy of Sciences, ul. Radzikowskiego 152, PL-31-342 
Krak{\'o}w, Poland}

\author{Antoni Szczurek}
\email{antoni.szczurek@ifj.edu.pl}
\affiliation{Faculty of Mathematics and Natural Sciences,
University of Rzesz\'ow, ul. Pigonia 1, PL-35-310 Rzesz\'ow, Poland}

\begin{abstract}
We calculate cross section for production of $\chi_c$ pairs
in proton-proton collisions. The cross section for the
$g g \to \chi_{c J_1} \chi_{c J_2}$ is considerably smaller
(especially for $\chi_{c 1} \chi_{c 1}$ final state) than
that obtained recently in the $k_T$-factorization approach.
We calculate therefore next-to-leading order contributions with
$\chi_c$ pair and one extra associated (mini-)jet.
We find these contributions to be much larger than those for the $2 \to 2$
contribution. Especially the emission of a leading gluon (carrying a large 
momentum fraction of one of the incoming gluons) are important.
These emissions in the $k_T$-factorization approach are absorbed into
the initial state unintegrated gluon distributions.
A smaller contribution to the cross section comes from  the production of central 
gluons emitted with rapidities between the $\chi_c$-mesons.
They do lead, however, to an enhancement of the $\chi_c$-pair production
at large rapidity distance between the mesons.
Our present study explains the size of the cross section for the $\chi_c$ pair 
production obtained previously in the $k_T$-factorization approach.
Several differential distributions are presented.
\end{abstract}

\pacs{12.38.Bx, 13.85.Ni, 14.40.Pq}
\maketitle
 
\section{Introduction}
The production of quarkonia in the nonrelativistic pQCD approach has
a long history. The production of $J/\psi$ is a good example, see
for example the review \cite{review}.
Using standard parameters of the $J/\psi$ wave functions the
lowest-order cross section in the color-singlet model 
is much below experimental data. 
Higher order corrections and/or color-octet contributions must be included to get closer to the data 
\cite{CampbellMaltoni:2007,GongLiWang:2010,Lansberg:2011}.
Furthermore, a large fraction of the prompt production originates from the radiative
decays of $P$-wave $\chi_c$ quarkonia.
Another efficient option is $k_T$-factorization approach \cite{k_T-fact} where already the lowest-order approach with unintegrated gluon distributions constructed 
following the prescription in Ref. \cite{Kimber:2001kmr}
gives reasonable results (see e.g. \cite{Baranov:2007dw,Baranov:2015yea,Baranov:2002cf,Kniehl:2006sk,Cisek:2017gno}).
In general, the inclusive cross section for $J/\psi$ (the same is true
for other quarkonia) grows with energy. 

In recent years also the production of $J/\psi$ pairs became accessible 
experimentally 
\cite{D0_jpsijpsi,LHCb_jpsijpsi_7TeV,Khachatryan:2014iia,
	Aaboud:2016fzt,Aaij:2016bqq}.
There is no yet sufficient understanding of the measured cross section. 
An important problem is the understanding of the contribution from single parton
scattering (SPS) and double parton scettaring (DPS) mechanisms.
Indeed, the importance of charm for the studies of double parton
scattering (DPS) has been stressed in \cite{Kom:2011bd,Luszczak:2011zp}.
Especially production of two $J/\psi$ mesons at large
rapidity difference is not well understood. The production of quarkonia
with large rapidity distance is often attributed to double parton
scattering mechanism for which the two partonic
processes are almost uncorrelated, in contrast to single parton
scattering mechanism where the correlation is encoded in relevant
matrix elements. 
In this region of phase space the DPS contribution to the cross section for 
different hard processes is well represented by the factorized ansatz:
\begin{equation}
\sigma({\rm DPS}, J/\psi J/\psi) = {1 \over 2} \, { \sigma^2({\rm SPS}, J/\psi) \over \sigma_{\rm eff}} \, .
\end{equation}
The so-called effective cross section $\sigma_{\rm eff}$ determines the normalization of the DPS contribution. 
A value of $\sigma_{ \rm eff} \approx 15 \, 
\rm{mb}$ was found from several phenomenological studies, see e.g. \cite{sigma_effective} or a 
table in Ref. \cite{Aaboud:2016fzt}.
In the case of $J/\psi$ pair production the cross section for large rapidity distances requires rather small values of $\sigma_{\rm eff} < 5 \, \rm{mb}$
\cite{D0_jpsijpsi,LHCb_jpsijpsi_7TeV,Khachatryan:2014iia,
	Aaboud:2016fzt,Aaij:2016bqq}.
Is the production of $J/\psi$ pairs different than for other
partonic processes? We do not see physical arguments to justify such a claim.

\begin{figure}[!h]
	\begin{minipage}{\textwidth}
		\centerline{\includegraphics[width=1.0\textwidth]{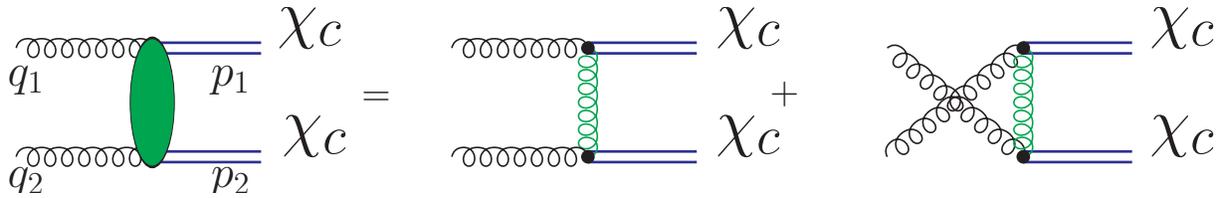}}
	\end{minipage}
	\caption{A diagrammatic representation of the leading order 
		mechanisms for $pp \to \chi_{c J_1} \chi_{c J_2}$ reaction.
	}
	\label{fig:diagram_chicchic}
\end{figure}

In Ref. \cite{Cisek:2017ikn} it was found that double $\chi_c$
production associated with radiative decays of both $\chi_c$ quarkonia 
leads to distributions quite similar to 
those from double parton scattering.
A rather sizeable cross section for $\chi_c$ pair production was
obtained from the $k_T$-factorization approach. 
Can we get a similar result within collinear-factorization approach? 
The 2 $\to$ 2 $g g \to \chi_c \chi_c$ processes were already calculated
long time ago \cite{Baranov:1997ph}.
We intend to calculate both $g g \to \chi_c \chi_c$ processes 
(see Fig.\ref{fig:diagram_chicchic}) as well as 2 $\to$ 3 processes 
(see Fig.\ref{fig:diagrams_chicchicg}). The recent calculation within 
$k_T$-factorization suggests that the 2 $\to$ 3 contributions may be
sizeable.

One would expect that the emission of a gluon in the central
rapidity region of the parton-level process
(see diagram (C) in Fig.\ref{fig:diagrams_chicchicg}) will enhance the
cross section at  large rapidity distances between the
$\chi_c$ mesons.
The contributions of  leading gluons, which carry a large longitudinal momentum fraction of one of the incoming gluons, (see diagrams (A) and (B) 
in Fig.\ref{fig:diagrams_chicchicg}) contain a contribution of minijets
produced at a large rapidity distance to the $\chi_c$-pair. Such contributions-
beyond the obvious collinear emissions-are included in the $k_T$-factorization
approach already in the lowest order. There these gluons are
absorbed into the initial state unintegrated gluon distribution.
The 2 $\to$ 3 precoesses were studied previously in the context of
quarkonium pair production for $p p \to J/\psi J/\psi g$ reaction 
\cite{Likhoded:2016zmk} and the corresponding cross section turned out to be
similar to the leading $p p \to J/\psi J/\psi$ contribution
and important in order to understand some correlation observables.

We will illustrate our calculations with several examples of
$\chi_c \chi_c$ pairs. Several differential distributions will be shown.

\section{Formalism}
\subsection{Parton-level amplitudes}

\begin{figure}[!h]
	\includegraphics[width=\textwidth]{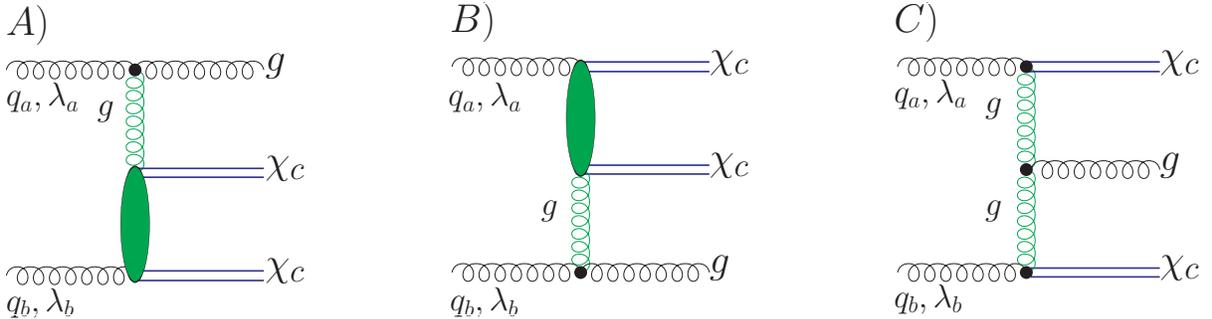}
	\caption{The lowest-order mechanisms for the  
		$\chi_{c J_1} \chi_{c J_2} g$ production in the high-energy kinematics described in the text.
	}
	\label{fig:diagrams_chicchicg}
\end{figure}

We are interested in three types of configurations in which a final state 
gluon is produced: 
firstly, the central production of a gluon $ gg \to \chi_{cJ} \, g \, \chi_{cJ}$
(diagram (C) in Fig. \ref{fig:diagrams_chicchicg}) and secondly the two configurations with leading gluons
(diagrams (A) and (B) in Fig. \ref{fig:diagrams_chicchicg}), where a gluon carries
the largest fraction of momentum of one of the incoming gluons.
The leading gluon minijet production is expected of importance for 
comparison to the $k_T$-factorization approach. 
This contribution is dominated by a kinematics, where
the gluon is emitted at large rapidity distance to the $\chi_c$ mesons.

A gauge invariant way to organize the calculation in this situation is the use
of vertices from the Lipatov effective action \cite{Lipatov:1996ts,Antonov:2004hh}. 

Let us introduce the four momenta of incoming protons, neglecting their masses,
\begin{eqnarray}
P_{1 \mu} = \sqrt{{s \over 2}} \, n^+_\mu \, , \, 
P_{2 \mu} = \sqrt{{s \over 2}} \, n^-_\mu \, ,
\end{eqnarray}
with the lightlike basis vectors
\begin{eqnarray}
n^\pm_\mu = {1 \over \sqrt{2}} (1,0,0,\pm 1) \, .
\end{eqnarray}
The incoming gluon momenta are
\begin{eqnarray}
q_a = q_a^+ n^+_\mu = x_1 P_{1\mu} \, , \, q_b = q_b^- n^-_\mu = x_2 P_{2 \mu} \, . 
\end{eqnarray}
The vertex for the ``upper'' leading gluon reads \cite{Lipatov:1996ts,Antonov:2004hh}
\begin{eqnarray}
n^{-\rho} \, \Gamma_{\mu \nu \rho}(q_a,p_1) = 2q^+_{a} g_{\mu \nu} + n^-_\mu (p_1 - 2q_a)_\nu + (q_a - 2p_1)_\mu n^-_\nu - {(p_1 - q_a)^2 \over q_a^+} n^-_\mu n^-_\nu \, ,\nonumber \\
\end{eqnarray}
while for the ``lower'' leading gluon we have
\begin{eqnarray}
n^{+\rho} \Gamma_{\mu \nu \rho}(q_b,p_2) = 2q^-_b g_{\mu \nu} + n^+_\mu (p_2 - 2q_b)_\nu + (q_b - 2p_2)_\mu n^+_\nu - {(p_2 - q_b)^2 \over q_b^-} n^+_\mu n^+_\nu \, .\nonumber \\
\end{eqnarray}
For the vertex of central gluon production (the ``Lipatov-vertex'') 
we introduce the momenta of fusing gluons
\begin{eqnarray}
q_{1\mu} = q_1^+ n^+_\mu + q^\perp_{1\mu}  \, , \,  q_{2\mu} = q_2^- n^-_\mu + q^\perp_{2\mu} \, , \, q_i^2 = (q^\perp_{i})^2 = - \vec{q}_{i\perp}^2 \, .  
\end{eqnarray} 
So that
\begin{eqnarray}
\Gamma_{\mu \rho \nu} (q_1,q_2) &=& n^-_\mu n^+_\nu C_\rho(q_1,q_2) \, , \, \nonumber \\
 C_\rho(q_1,q_2) &=& (q_1^+ + {q_1^2 \over q^-_2 })  n^+_\mu -(q_2^- + {q_2^2 \over q^+_1} ) n^-_\mu + (q_2 - q_1)^\perp_\mu \, . 
\end{eqnarray}
We also need the $g^* g^* \to \chi_{cJ}$ vertices.
We write them in the form
\begin{eqnarray}
V^{ab}_{\mu \nu}(J,J_z;q_1,q_2) = -i 4 \pi \alpha_S \delta^{ab} {2 R'(0) \over \sqrt{\pi N_c M^3}}
	\sqrt{3} \, \cdot T_{\mu \nu}(J,J_z;q_1,q_2) \, .
\end{eqnarray}
The explicit form of the tensors $T_{\mu \nu}$ are found in Ref.\cite{Cisek:2017ikn}.
Above $a,b$ are the color indices of incoming gluons, $N_c =3$ is the number of colors, and $M$ is the mass of the $\chi_c$ meson.
For $J=1$ and $J=2$ states, the tensors have the form
\begin{eqnarray}
T_{\mu \nu}(1,J_z;q_1,q_2) &=& T_{\mu \nu \alpha}(1;q_1,q_2 ) \varepsilon^{\alpha*}(J_z,q_1+q_2) \, , \nonumber \\
T_{\mu \nu}(2,J_z;q_1,q_2) &=& T_{\mu \nu \alpha \beta}(2;q_1,q_2 ) \varepsilon^{\alpha \beta*}(J_z,q_1+q_2) \, ,
\end{eqnarray}
where $\varepsilon_\mu(J_z,p), \varepsilon_{\mu\nu}(J_z,p)$ is the polarization vector/tensor for the meson with momentum $p$.
The derivative of the radial wave function at the origin is related to the $\gamma \gamma$-decay width as
\begin{eqnarray}
\Gamma(\chi_{c0} \to \gamma \gamma) = {27 e_c^4 \alpha_{\rm em}^2 \over m_c^4} |R'(0)|^2 \, .
\end{eqnarray}
We use the value $|R'(0)|^2 = 0.042 \, \rm{GeV}^2$.
We can now construct all the $2 \to 3$ amplitudes of interest from
the above tensors.
The amplitude for $gg \to \chi_{cJ_1} g \chi_{cJ_2}$ with a central gluon 
reads:
\begin{eqnarray}
{\cal M}_C = i g_S f_{a'b'c} \, V^{aa'}_1(q_a,p_1) {1 \over t_1} \, C^\rho(q_a-p_1,q_b-p_2) \varepsilon^*_\rho(\lambda_g,p_g) \, {1 \over t_2}  V^{bb'}_2(q_b,p_2) \, ,
\label{eq:central_amplitude}
\end{eqnarray}
where
\begin{eqnarray}
 V^{aa'}_1(q_a,p_1) &=& \varepsilon^\mu(\lambda_a,q_a) V^{aa'}_{\mu \mu'}(J_1,J_{1z};q_a,p_1-q_a) n^{-\mu'}
 \, , \nonumber \\
 V^{b b'}_2(q_b,p_2) &=& \varepsilon^\nu(\lambda_b,q_b) V^{bb'}_{\nu' \nu}(J_2,J_{2z};q_b,p_2-q_b) n^{+\nu'}	 \, .
\end{eqnarray}
The amplitude for the final state with the leading gluons in the fragmentation region of gluon $q_a$ or $q_b$ can be written in terms of
the (half-) off-shell amplitude for the $g^* g \to \chi_{c1} \chi_{c2}$
process.
The $2\to 2$ amplitude is obtained from
\begin{eqnarray}
{\cal M}^{ab}_{\mu \nu}(q_a,q_b;p_1,p_2) &=& V^{aa'}_{\mu \mu'}(J_1,J_{z1};q_a,p_1-q_a) {-g^{\mu'\nu'} \delta^{a'b'} \over \hat t}  V^{bb'}_{\nu' \nu}(J_1,J_{z1};p_2-q_b,q_b) \nonumber \\
&&+  V^{bb'}_{\nu \nu'}(J_1,J_{z1};q_b,p_1-q_b) {-g^{\mu'\nu'} \delta^{a'b'} \over \hat u}  V^{aa'}_{\mu' \mu}(J_1,J_{z1};p_1-q_a,q_a) \, . \nonumber \\
\label{eq:2_to_2}
\end{eqnarray} 
Here the Mandelstam variables are 
\begin{equation}
\hat t = (p_1-q_a)^2 = (p_2-q_b)^2\, , \, 
\hat u = (p_1-q_b)^2 = (p_2 - q_a)^2 \, . 
\end{equation}
The amplitude of Eq. (\ref{eq:2_to_2}) enters the $2 \to 3$ amplitudes as follows:
\begin{eqnarray}
{\cal M}_A &=& i g_S f_{ab'c} \varepsilon^\mu(\lambda_a,q_a) \Gamma_{\mu \nu \rho}(q_a,p_g) n^{-\rho} \varepsilon^{\nu*}(\lambda_g,p_g)
{1 \over t_1} n^{+\mu'} {\cal M}^{b'b}_{\mu' \nu'}(p_g-q_a,q_b;p_1,p_2) \varepsilon^{\nu'}(\lambda_b,q_b) \nonumber \\
&=& i g_S f_{ab'c} \,  2q_a^+ \delta_{\lambda_a \lambda_g} {1 \over t_1} \,  n^{+\mu'} \varepsilon^{\nu'}(\lambda_b,q_b)   {\cal M}^{b'b}_{\mu' \nu'}(p_g-q_a,q_b;p_1,p_2) \, ,
\end{eqnarray}
and likewise
\begin{eqnarray}
{\cal M}_B &=& i g_S f_{a'bc} n^{-\nu'} \varepsilon^{\mu'}(\lambda_a,q_a)   {\cal M}^{a a'}_{\mu' \nu'}(q_a,p_g-q_b;p_1,p_2) \, { 1\over t_2} 
\varepsilon^\mu(\lambda_b,q_b) \Gamma_{\mu \nu \rho}(q_b,p_g) n^{+\rho} \varepsilon^{\nu*}(\lambda_g,p_g) \nonumber \\
&=& i g_S f_{a'bc} n^{-\nu'} \varepsilon^{\mu'}(\lambda_a,q_a)  
 {\cal M}^{a a'}_{\mu' \nu'}(q_a,p_g-q_a;p_1,p_2) 
{ 1 \over t_2}
\, 2 q_b^- \delta_{\lambda_b \lambda_g} \, .
\end{eqnarray}
We close this section with a brief comment on the gluon exchanges in the crossed channel. The $t$-channel gluons explicitly depicted in Fig.\ref{fig:diagrams_chicchicg} are taken in the respective high-energy limit
- they correspond to the reggeized gluons of the effective action \cite{Antonov:2004hh,Lipatov:1996ts}. For the gluon exchanges in the blobs
of diagrams (A) and (B) of Fig.\ref{fig:diagrams_chicchicg} we checked that the approximation of reggeized gluon exchange in the $gg \to \chi_c \chi_c$ subprocess becomes
a good approximation at a rapidity distance between $\chi_c$'s of 
$\Delta y \gtrapprox 3$. In the numerical calculations, we use the 
full gluon propagator in Feynman-gauge.
We note that the interference between $t$- and $u$-channel amplitudes is negligible and confined to a very narrow interval around $\Delta y \sim 0$. 
\subsection{Parton-level cross sections}
Let us now have a look at the parton-level cross section in order to understand better the kinematics and possible singularities in the integration
over phase space.
The $2 \to 3$ parton-level cross sections are obtained from
\begin{eqnarray}
d \sigma = {1 \over 4 q_a^+ q_b^-} \, | \overline{{\cal M}_i}|^2 \, d\Phi(q_a+q_b; p_1,p_2,p_g) \, ,
\end{eqnarray}
where $i = A,B,C$, and there is no interference between the diagrams
of Fig.\ref{fig:diagrams_chicchicg}.
Let us start from the production of a leading gluon along the direction of 
incoming gluon $a$, described by amplitude ${\cal M}_A$.
Here, following the rules of the high-energy limit, the four-momentum $q_1 \equiv p_g - q_a$ of the exchanged 
gluon enters the $2 \to 2$ amplitude in the form
\begin{eqnarray}
q_{1\mu} = q^+_1 n^+_\mu + q^\perp_{1\mu} \equiv z_1 q^+_a n^+_\mu +   q_{1\perp } e^\perp_{\mu} \, .
\end{eqnarray}
We can now use the Ward-identity, to write
\begin{eqnarray}
n^{+\mu} \varepsilon^{\nu}(\lambda_b,q_b)   {\cal M}^{b'b}_{\mu \nu}(q_1,q_b;p_1,p_2) &=& {q_{1\perp} \over q^+_1} e^{\perp \mu} \varepsilon^{\nu}(\lambda_b,q_b)   
{\cal M}^{b'b}_{\mu \nu}(q_1,q_b;p_1,p_2) 
\nonumber \\
&\equiv& {q_{1\perp} \over q^+_1} 
{\cal M}(2 \to 2)\, . 
\end{eqnarray}
Then, the $2 \to 3$ cross section takes the simple form
\begin{eqnarray}
d \sigma(2 \to 3) &=& {2 C_A \alpha_S \over \pi} {q^2_{1\perp} \over t_1^2} {d^2  \vec{q}_{1\perp} \over \pi} {dz_1 \over z_1 (1-z_1)} \, {1 \over 4 q_1^+ q_b^-}  |\overline{{\cal M}(2 \to 2)}|^2 \, d\Phi(q_1 + q_b; p_1,p_2)  
\nonumber \\
&=& {2 C_A \alpha_S \over \pi} {d^2  \vec{q}_{1\perp} \over \pi  q^2_{1 \perp}}
{dz_1 \over z_1} \, \cdot {1 \over 2 q_1^+ q_b^-}  |\overline{{\cal M}(2 \to 2)}|^2 \, d\Phi(q_1 + q_b; p_1,p_2)  \, .
\end{eqnarray}
Here one would recognized the factorization in the unintegrated gluon distribution in a gluon
\begin{eqnarray}
{z d n_{g/g}(z,\vec{q}_\perp) \over dz d \log q_\perp^2 } = {2 C_A \alpha_S \over \pi} \, ,
\end{eqnarray}
and the off-shell cross section for the process $g^* g \to \chi_c \chi_c$.
The off-shell cross section will provide us with a scale $\mu^2 \sim M^2_\perp$, so that for $q_{1 \perp}^2 \ll \mu^2$ we can neglect the off-shellness of gluon $q_1$ and only the on shell cross section $gg \to \chi_c \chi_c$ enters.
The parton-level cross section then consists of two parts:
\begin{eqnarray}
d \sigma(2 \to 3) = {2 C_A \alpha_S \over \pi} 
 \int^{\mu^2} {d  q^2_{1\perp} \over   q^2_{1 \perp}}
{dz_1 \over z_1} \, d\sigma(2 \to 2)  + {2 C_A \alpha_S \over \pi} 
\int_{\mu^2} {d^2  \vec{q}_{1\perp} \over \pi  q^2_{1 \perp}}
{dz_1 \over z_1} d\sigma(2 \to 2; \vec{q}_{1\perp} ) \nonumber \\
\end{eqnarray} 
Here the first piece contains the infrared divergent integral 
$\int^{\mu^2} d  q^2_{1\perp} /   q^2_{1 \perp}$, which is of course
just the collinear logarithm in the $g \to gg$ splitting. In a complete
NLO calculation of the inclusive $\chi_c \chi_c$ the collinear logarithm within some
factorization scheme would be absorbed into the evolution
of the gluon distribution of one of the protons. The contribution from hard
$q^2_{\perp1} > \mu^2$ is a genuine NLO contribution.
In our numerical calculations we will simply show the $2 \to 3$ cross section
with a lower cutoff on the transverse momentum of the produced gluon (mini-)jet,
$p_{g \perp} > p^{\rm cut}_{g \perp} \sim 1 \, \rm{GeV}$.

Let us now come to the contribution from production of a central gluon in
the $gg \to \chi_c g \chi_c$ process.
We write the parton-level cross section differential in the gluon rapdity $y_g$ and
the transverse momenta of $\chi_c$ mesons $\vec{p}_{1,2\perp}$.
\begin{eqnarray}
{d \sigma (gg \to \chi_c g \chi_c)} = { 1 \over 256 \pi^5 \hat s^2} 
|\overline{{\cal M }_C}|^2 \, dy_g d^2\vec{p}_{1\perp} d^2\vec{p}_{2\perp} \, .
\end{eqnarray}
The square of the amplitude ${\cal M}_C$ of Eq.(\ref{eq:central_amplitude}) 
can be written in the usual impact factor representation
\begin{eqnarray}
\overline{|{\cal M}_C|^2} &=& {N_c \over N_c^2 -1} 16 \pi \alpha_S  \, 
I_1( \vec{p}_{1\perp}) {\hat s^2 \over (\vec{p}_{1\perp} + \vec{p}_{2\perp} )^2}
I_2( \vec{p}_{2\perp}) \, \nonumber \\
&=& {16 \pi^3 \hat{s}^2 \over N_c^2 -1} I_1( \vec{p}_{1\perp}) 
{\cal K}_r(\vec{p}_{1\perp}, - \vec{p}_{2\perp}) 
I_2( \vec{p}_{2\perp}) \, . \nonumber \\
\end{eqnarray}
Here ${\cal K}_r$ is the real-emission part of the BFKL-kernel \cite{BFKL}
\begin{eqnarray}
{\cal K}_r(\vec{p}_{1\perp}, - \vec{p}_{2\perp}) = {C_A \alpha_S \over \pi^2}  {1 \over (\vec{p}_{1\perp} + \vec{p}_{2\perp} )^2} \, .
\end{eqnarray}
Notice that the integral over the gluon rapidity is proportional
to $Y = \log(\hat s / M^2)$, so that the $2 \to 3$ cross section will be
\begin{eqnarray}
d\sigma(2 \to 3) = {Y \over 16 \pi^2 (N_c^2 -1)} \, I_1(\vec{p}_{1\perp})
{\cal K}_r(\vec{p}_{1\perp}, - \vec{p}_{2\perp}) I_2(\vec{p}_{2\perp}) d^2\vec{p}_{1\perp} d^2\vec{p}_{2\perp} \, .
\end{eqnarray}
Here we again have an infrared singularity when $\vec{p}_{g\perp} = - \vec{p}_{1\perp} - \vec{p}_{2\perp} \to 0$. 
This is of course just the back-to-back region of the $2 \to 2$ process.
The differential cross section of the Born-level $2 \to 2$ cross section can be expressed in terms of the
same impact factors and reads
\begin{eqnarray}
d \sigma^{(0)}(2 \to 2) = {1 \over 16 \pi^2 (N_c^2 -1)} I_1(\vec{p}_{1\perp})
\delta^{(2)} (\vec{p}_{1\perp}+ \vec{p}_{2\perp}) I_2(\vec{p}_{2\perp}) d^2\vec{p}_{1\perp} d^2\vec{p}_{2\perp} \, .
\end{eqnarray}
To the leading order in $\alpha_S Y$, the virtual correction to the $2 \to 2$ process can be easily calculated using the gluon reggeization property, which
amounts to the replacement of the gluon propagator by
\begin{eqnarray}
{1 \over q^2} \to {1 \over q^2} \, \exp[\omega(\vec{q}_\perp) Y] \, ,
\end{eqnarray}
where 
\begin{eqnarray}
\omega(\vec{q}_{\perp}) = -{\alpha_S N_c \over 4 \pi^2} \int d^2\vec{Q}_\perp  { \vec{q}^2_\perp \over \vec{Q}^2_\perp (\vec{Q}_\perp - \vec{q}_\perp)^2} \, .
\end{eqnarray}
Expanding the Regge-propagator to the first order,
we obtain the $2 \to 2$ process cross section as
\begin{eqnarray}
d \sigma( 2 \to 2) &=& d \sigma^{(0)}(2 \to 2) \nonumber \\
&+& 
{Y \over 16 \pi^2 (N_c^2 -1)} I_1(\vec{p}_{1\perp})
\delta^{(2)} (\vec{p}_{1\perp}+ \vec{p}_{2\perp}) 2 \omega(\vec{p}_{1\perp}) I_2(\vec{p}_{2\perp}) d^2\vec{p}_{1\perp} d^2\vec{p}_{2\perp} \, . \nonumber \\
\end{eqnarray} 
Then, the inclusive cross section for the production of $\chi_c$-pairs
becomes
\begin{eqnarray}
d \sigma(gg \to \chi_c \chi_c X) = d\sigma^{(0)}  + { Y \over 16 \pi^2 (N_c^2 -1)} 
I_1(\vec{p}_{1\perp}) {\cal K}_{\rm BFKL} (\vec{p}_{1\perp}, - \vec{p}_{2\perp}) I_2(\vec{p}_{2\perp}) d^2\vec{p}_{1\perp} d^2\vec{p}_{2\perp} \, .\nonumber \\
\end{eqnarray}
Here ${\cal K}_{\rm BFKL}$ is the leading-order in $\alpha_S Y$ 
BFKL kernel
\begin{eqnarray}
{\cal K}_{\rm BFKL} (\vec{p}_{1\perp}, - \vec{p}_{2\perp})
&=& {\cal K}_{r} (\vec{p}_{1\perp}, - \vec{p}_{2\perp}) + {\cal K}_{\rm v} (\vec{p}_{1\perp}, - \vec{p}_{2\perp}) \nonumber \\
&=& {\alpha_S N_c \over \pi^2} \Big( {1 \over (\vec{q}_{1\perp} + \vec{q}_{2\perp})^2 } - \delta^{(2)}(\vec{q}_{1\perp} +\vec{q}_{2\perp}) {1 \over 2} \int d^2 {\vec Q}_\perp  
{\vec{q}^2_\perp \over \vec{Q}^2_\perp (\vec{Q}_\perp - \vec{q}_\perp)^2} \Big) \, . \nonumber \\
\end{eqnarray}
We cannot absorb the infrared divergencies into the initial state parton distributions in this case.
However, for the sufficiently inclusive, say over soft gluon radiation, cross section, the infrared divergencies in the real and virtual part of the BFKL-kernel will cancel.
Notice that this mechanism resembles in many respects the Mueller-Navelet dijet production \cite{Mueller:1986ey}, with the $\chi_c$ playing the role of the jets. 
However, for this case more involved calculations including a full BFKL-resummation
have been performed in recent years \cite{Ducloue:2013bva,Caporale:2014gpa}.
As in the case for production of leading gluons, we will in
our numerical calculations show the contribution from the
$\chi_c \chi_c g$ final state with a lower cutoff on the
transverse momentum of the gluon  
$p_{g \perp} = |\vec{q}_{1 \perp} + \vec{q}_{2 \perp} |
> p^{\rm cut}_{g \perp} \sim 1 \, \rm{GeV}$.
\subsection{Hadron-level cross sections}
We now come to the hadron-level cross sections.
Below $s$ is the proton-proton center-of-mass energy squared.
The inclusive production of $\chi_c$-pairs from the $2 \to 2$ process is obtained from
\begin{eqnarray}
\displaystyle
d\sigma &=& x_1g(x_1,\mu^2) x_2g(x_2,\mu^2) {1 \over 16 \pi (x_1 x_2 s)^2} | \overline{\mathcal{M}(2 \to 2)} |^2 dy_1 dy_2 
d^2 \vec{p}_{1\perp} d^2 \vec{p}_{2\perp} 
\delta^{(2)}(\vec{p}_{1 \perp} + \vec{p}_{2\perp}) \, , 
\nonumber \\
\end{eqnarray}
with $p_T = |\vec{p}_{1\perp}|$, and
\begin{eqnarray}
x_1 = \sqrt{ {M^2 + p_T^2 \over  s}}\Big( e^{y_1} + e^{y_2} \Big) \, , \, 
x_2 =  \sqrt{ {M^2 + p_T^2 \over  s}}\Big( e^{-y_1} + e^{-y_2} \Big) \, .
\end{eqnarray}
The cross section for the $ 2\to 3$ processes is calculated
from:
\begin{eqnarray}
d \sigma &=& x_1 g(x_1,\mu^2) x_2 g(x_2,\mu^2) {1 \over 256 \pi^5  (x_1 x_2 s)^2} | \overline{\mathcal{M}(2 \to 3)} |^2 
\nonumber \\
&&\times
dy_1 dy_2 dy_g 
d^2 \vec{p}_{1\perp} d^2\vec{p}_{2\perp} 
d^2 \vec{p}_{g\perp} 		
	\delta^{(2)}(\vec{p}_{1 \perp} + \vec{p}_{2\perp} + \vec{p}_{g\perp}) 
\end{eqnarray}
with 
\begin{eqnarray}
\displaystyle
x_{1}&=&\frac{m_{1\perp}}{\sqrt{s}}e^{y_{1}} + \frac{m_{2\perp}}{\sqrt{s}}e^{y_{2}}+\frac{p_{g\perp}}{\sqrt{s}}e^{y_{g}}\, ,\\
x_{2}&=&\frac{m_{1\perp}}{\sqrt{s}}e^{-y_{1}} + \frac{m_{2\perp}}{\sqrt{s}}e^{-y_{2}}+\frac{p_{g\perp}}{\sqrt{s}}e^{-y_{g}} \, ,
\end{eqnarray}
where $m_{i\perp} = \sqrt{M^2 + p_{i \perp}^2}$ and $y_{1,2}$ are the cm-rapidities of mesons.
We take as the factorization scale $\mu^2 = \hat s = x_1 x_2 s$.
For the case of identical $\chi_c$-mesons in the final state 
all of the cross sections must be multiplied by $1/2$.

\section{Numerical Results}

\subsection{Parton level processes}

In this subsection we show two examples of rapidity distributions
on the parton level for the process (C) in Fig.\ref{fig:diagrams_chicchicg}.

\begin{figure}[h!]
	\includegraphics[width = 8cm]{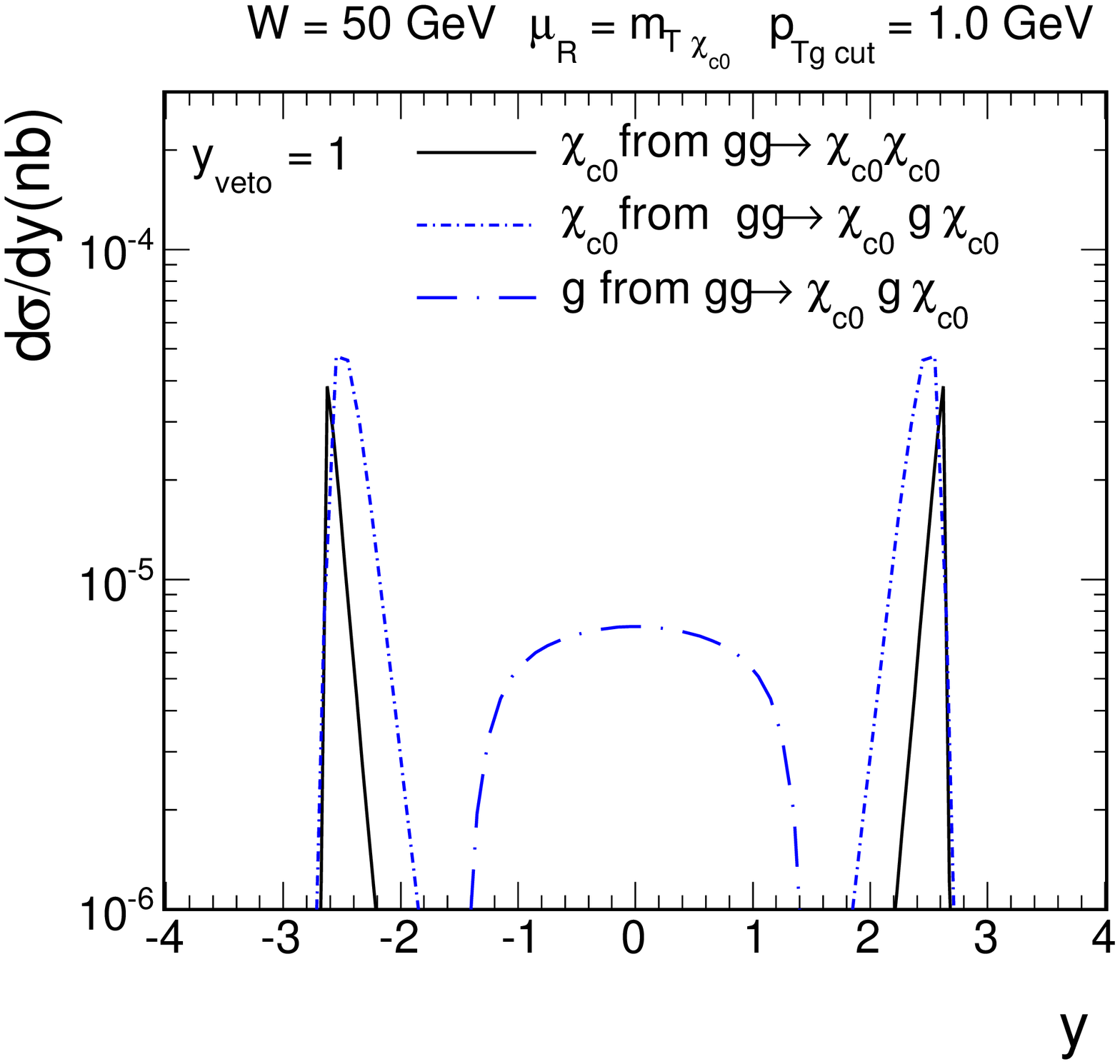}
    \includegraphics[width = 8cm]{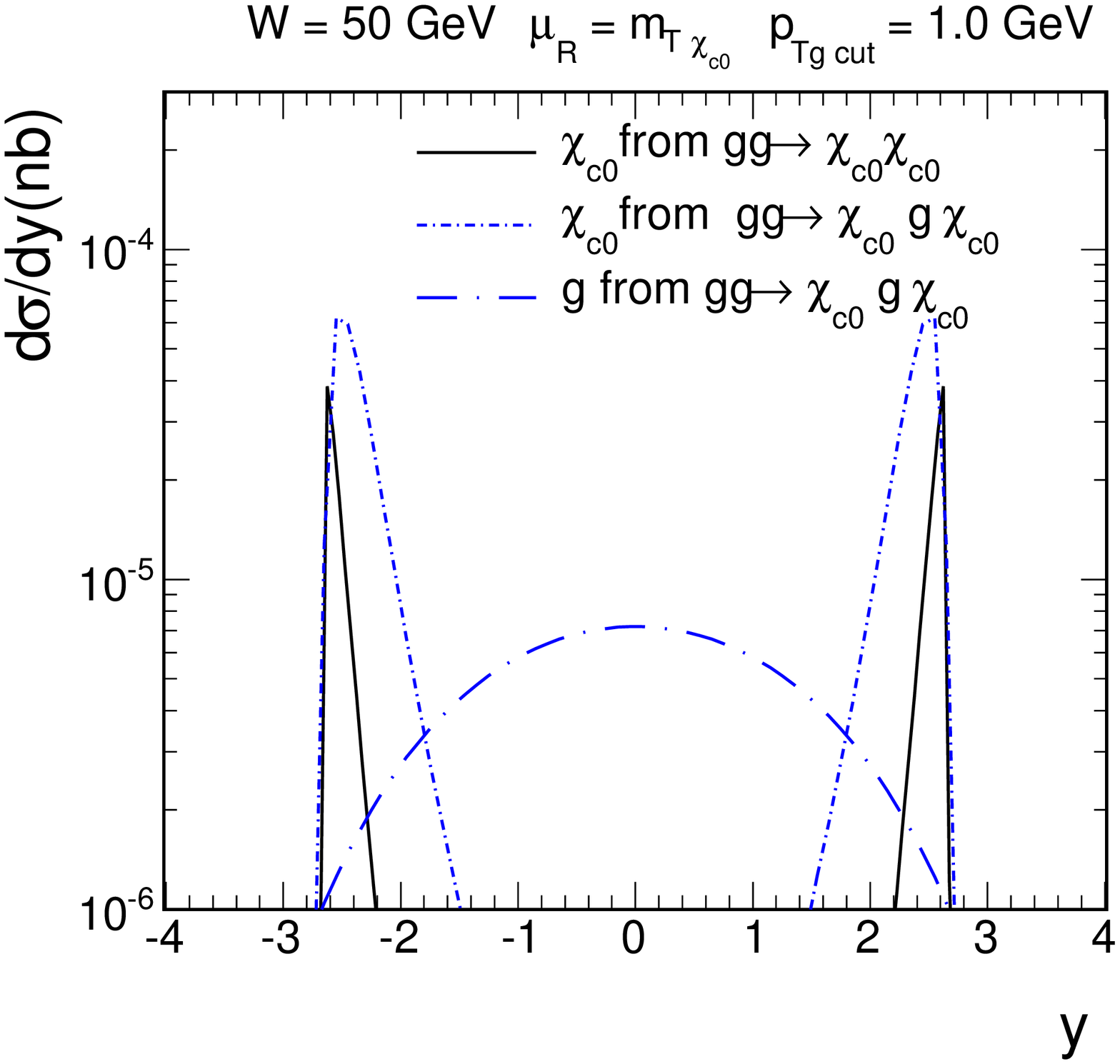}
	\caption{Differential cross section for the processes from Fig.\ref{fig:diagram_chicchic}
		and Fig.\ref{fig:diagrams_chicchicg} (C) at the parton level, where the energy in the center 
		of mass of two gluons was fixed for $W = 50\:\rm{GeV}$. The left panel is for extra rapidity veto and the right panel without the extra condition.}
	\label{fig:chic0chic0_partonic}
\end{figure}

In Fig.\ref{fig:chic0chic0_partonic} we show distribution in rapidity 
for the $g g \to \chi_{c 0} g \chi_{c 0}$. Here the center of mass energy
has been fixed at $W$ = 50 GeV. The two $\chi_{c 0}$ mesons are produced
in forward and backward directions while gluon at midrapidity
in the partonic center-of-mass system.
For comparison we show also rapidity distributions of $\chi_{c 0}$
mesons from the $g g \to \chi_{c 0} \chi_{c 0}$ process (solid line).

\begin{figure}[h!]
	\includegraphics[width = 8cm]{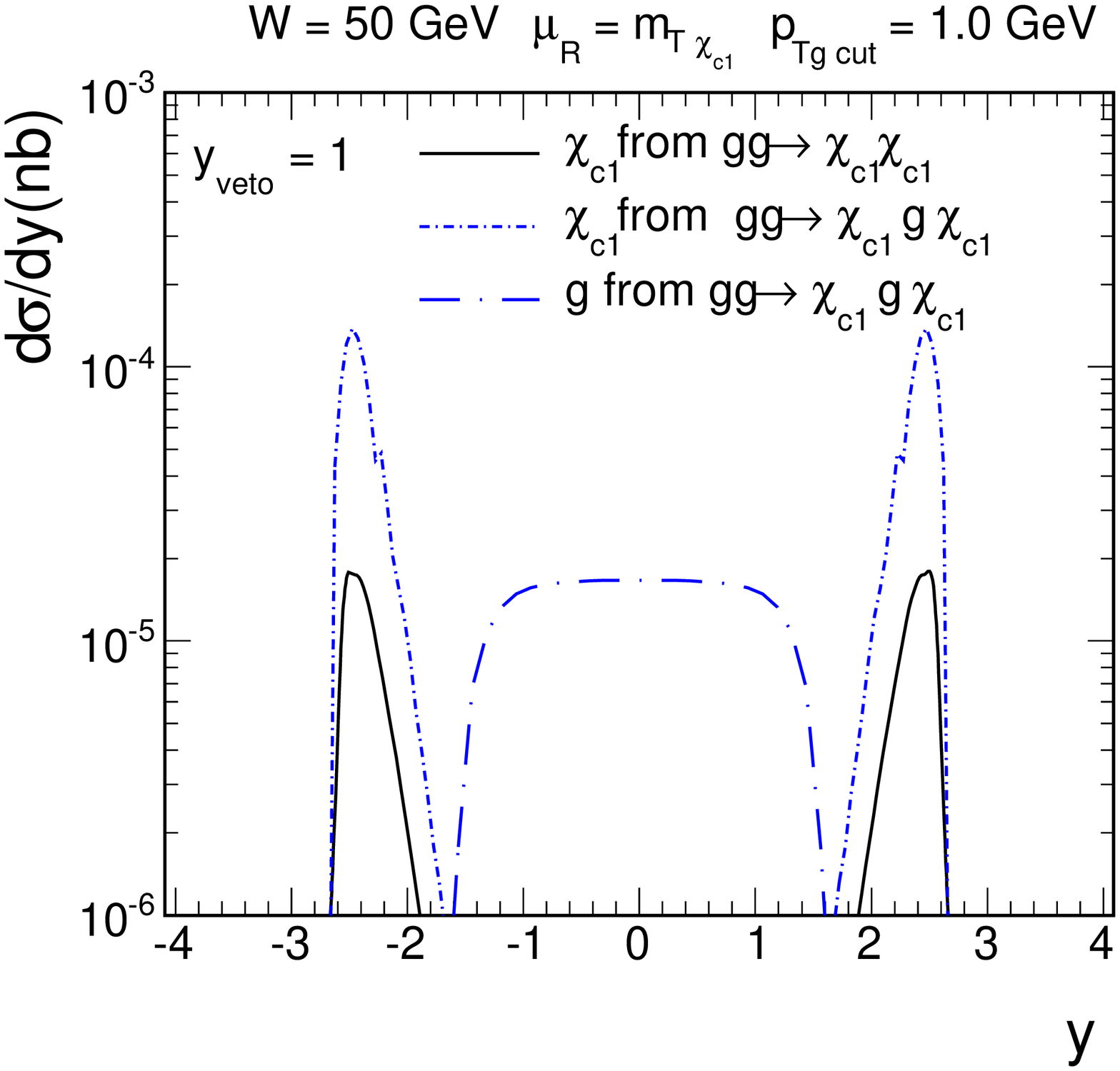}
	\includegraphics[width = 8cm]{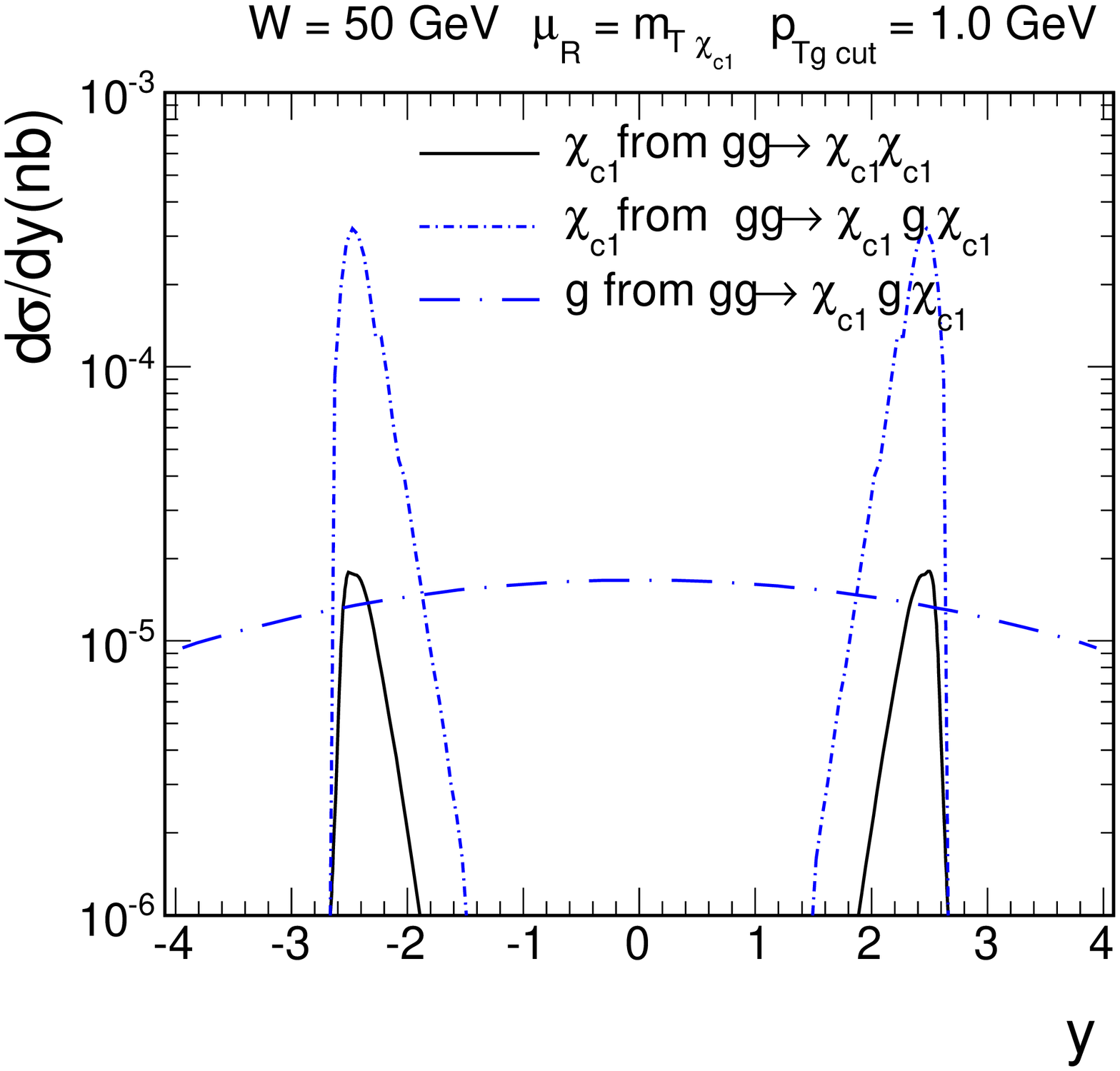}
	\caption{Differential cross section for the processes from Fig.\ref{fig:diagram_chicchic}
		and Fig.\ref{fig:diagrams_chicchicg} (C) at the parton level, where the energy in the center 
		of mass of two gluons was fixed for $W = 50\:\rm{GeV}$. The left panel is for extra rapidity veto and the right             panel without the extra condition.}
	\label{fig:chic1chic1_partonic}
\end{figure}

Similar distribution for the $g g \to \chi_{c 1} g \chi_{c 1}$ process
is shown in Fig.\ref{fig:chic1chic1_partonic}.
The situation is similar as for the $\chi_{c 0} \chi_{c 0}$ pair production. 
However, the $2 \to 3$ contribution here is relatively enhanced compared
to the $2 \to 2$ one (solid lines). 
In each of the $gg\rightarrow\chi_{c1}$ vertices in the $2\rightarrow2$ process only one gluon
is off-mass-shell, whereas in the $2\rightarrow3$ process in one of the vertices both 
gluons are off-mass-shell. The vertex $g^*g^*\rightarrow\chi_{c1}$ strongly depends on virtualities
of the gluons. We remind that when gluons are on-mass-shell the vertex vanishes (Landau-Yang theorem \cite{Landau-Yang}).

To ensure validity of the effective Regge action (applicability of the
Lipatov-vertex) one should ensure that the gluon is produced at a distance
of at least $y_{\rm veto} \sim 1$ from the mesons. We therefore show in the left panels
of Figs. \ref{fig:chic0chic0_partonic},\ref{fig:chic1chic1_partonic} the
result obtained for $y_{\rm veto} = 1$ and in the right panels the result
without a rapidity veto. Interestingly, for the $\chi_{c 0}$ case, the 
gluon is automatically produced centrally, while for the case of $\chi_{c 1}$
production the rapidity veto is important to exclude contributions from
non-central kinematics.

\newpage
\subsection{Hadron level cross sections}


The integrated cross sections (full phase space) for different components are shown
in Table \ref{tab:integrands} for $\sqrt{s}= 8\,\rm{TeV}$. We restrict ourselves to the case
of ``identical'' pairs, i.e. $\chi_{c0} \chi_{c0}$, $\chi_{c1} \chi_{c1}$, or $\chi_{c2} \chi_{c2}$. We see, that the
cross sections for the $2 \to 2$ processes are consistently lower
than the ones obtained in the $k_T$-factorization approach in Ref.
\cite{Cisek:2017ikn}.

\begin{figure}[!h]
\includegraphics[width=5.3cm]{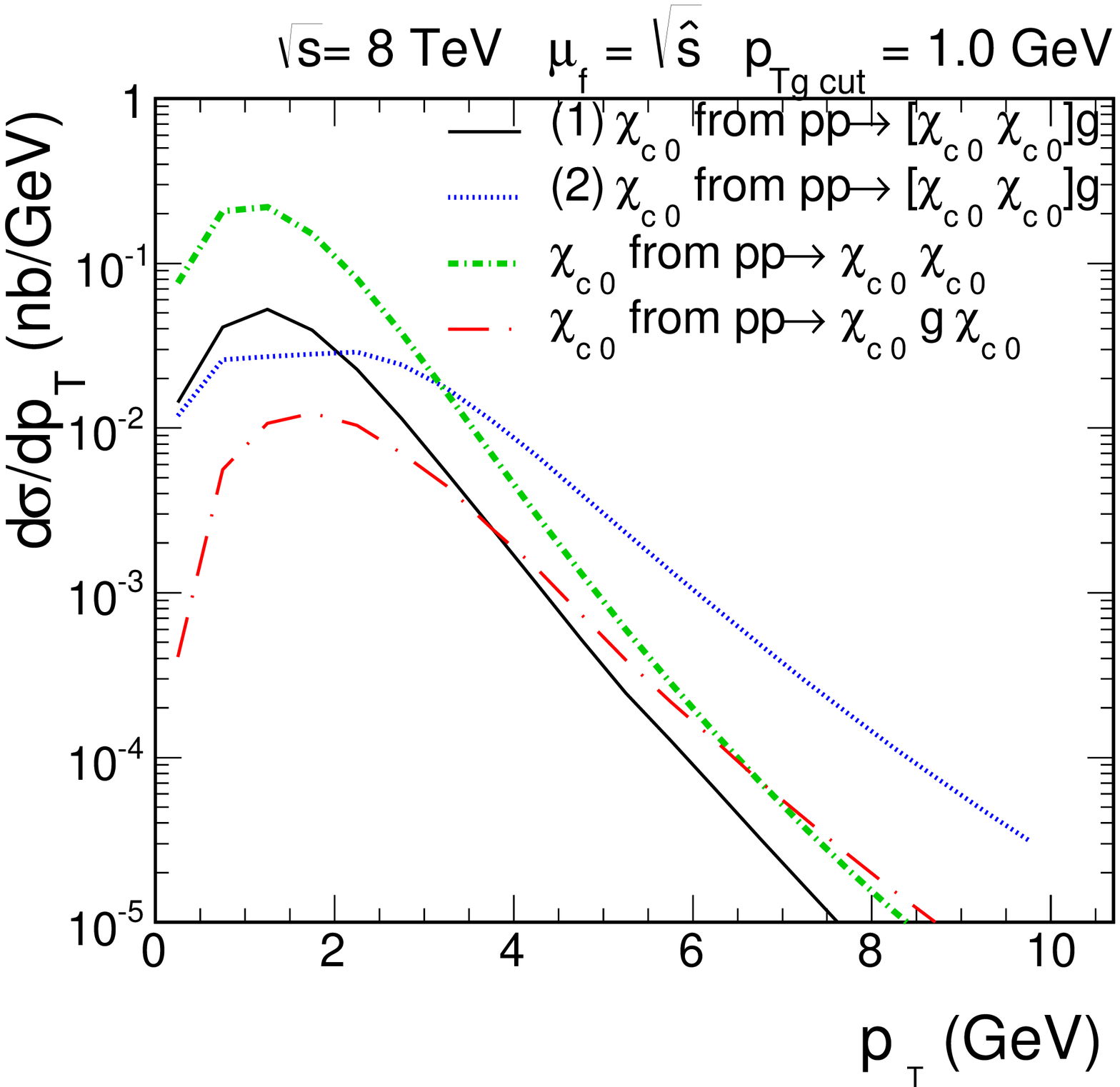}
\includegraphics[width=5.3cm]{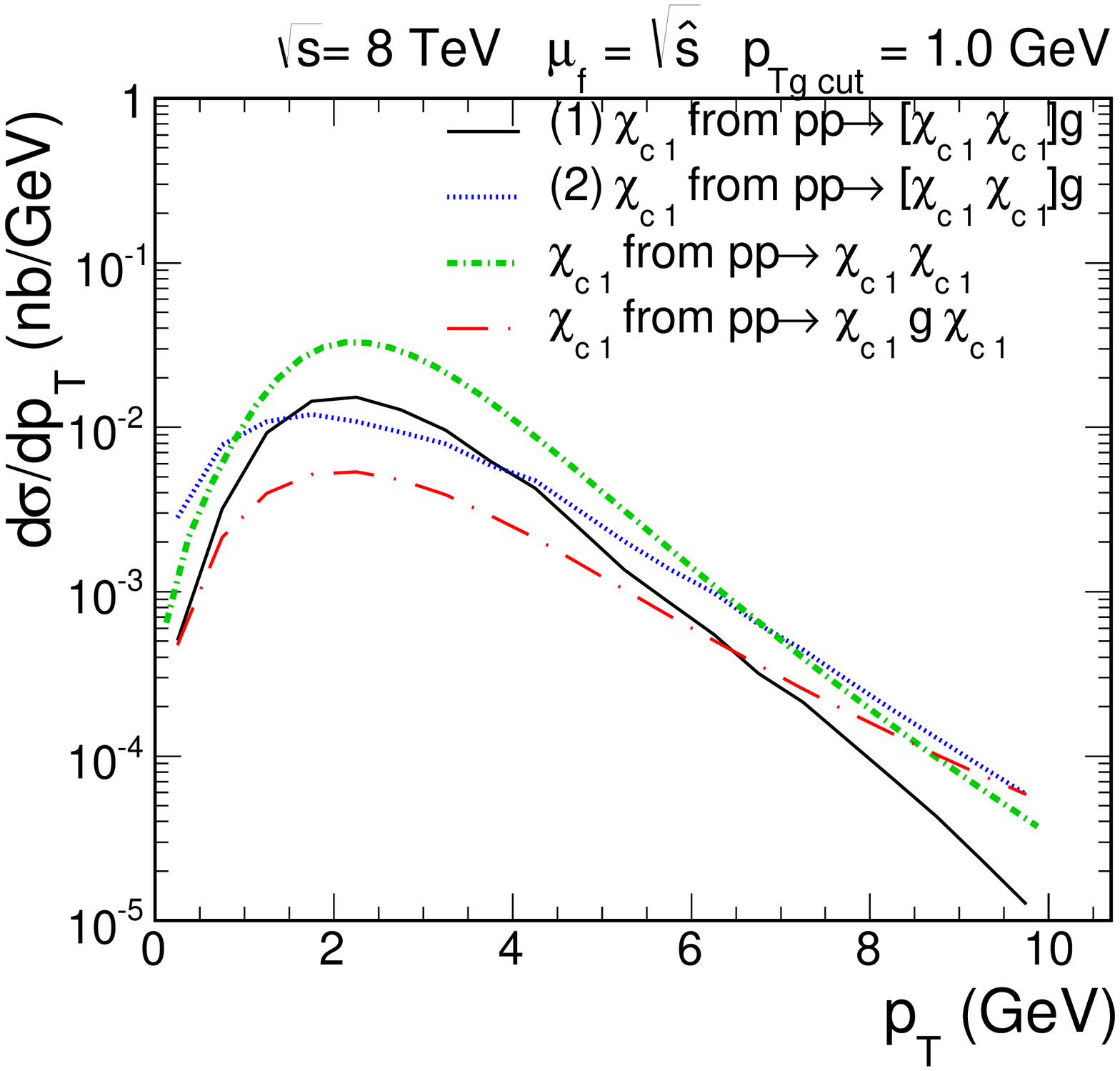}
\includegraphics[width=5.3cm]{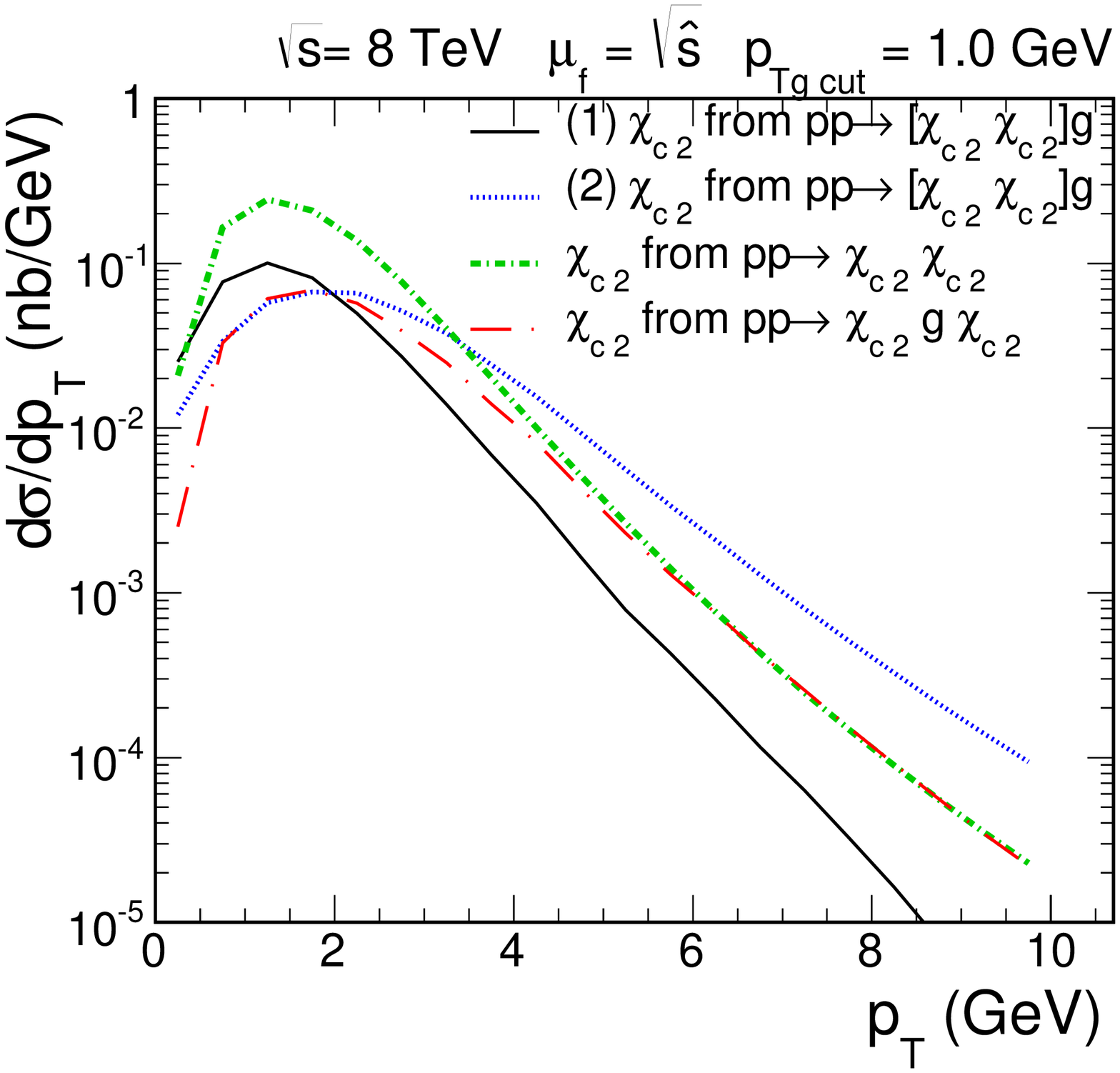}
   \caption{Transverse momenta distributions of one of the $\chi_c$
for the $p p \to \chi_{c J} \chi_{c J}$ 
and $pp \to \chi_{c J} \chi_{c J} g$ reaction for $\sqrt{s} = 8\:TeV$.
 }
\label{fig:dsig_dpt}
\end{figure}

\begin{table}[h!]
	\caption{Values of total cross sections for particular processes for $\sqrt{s}=8\,\rm{TeV}$.}
	\label{tab:integrands}
	
	\begin{tabular}{|l|c|l|c|l|c|} 
		\hline
		$\chi_{c 2}$ & $\sigma_{\rm total} $ & $\chi_{c 1}$ & $\sigma_{\rm total} $& $\chi_{c 0}$ & $\sigma_{\rm total} $\\
		\hline
		$pp\rightarrow\chi_{c 2} \chi_{c 2}$ & $0.62 \, \rm{nb}$ & $pp\rightarrow\chi_{c 1} \chi_{c 1}$ & $8.60\cdot 10^{-2} \, \rm{nb}$ & $pp\rightarrow\chi_{c 0} \chi_{c 0}$ & $0.40 \, \rm{nb}$ \\ 
		\hline
		$pp\rightarrow[\chi_{c 2} \chi_{c 2}]g$ & $0.19\, \rm{nb} \times 2$ & $pp\rightarrow[\chi_{c 1} \chi_{c 1}]g$ & $4.07\cdot 10^{-2}\, \rm{nb}\times 2$ & $pp\rightarrow[\chi_{c 0} \chi_{c 0}]g$ & $0.10 \, \rm{nb}\times 2$ \\
		\hline
		$pp\rightarrow \chi_{c 2} g \chi_{c 2}$ & $0.16\, \rm{nb}$ & $pp\rightarrow \chi_{c 1} g \chi_{c 1}$ & $ 1.78\cdot 10^{-2} \, \rm{nb}$ & $pp\rightarrow \chi_{c 0} g \chi_{c 0}$ & $0.03 \, \rm{nb}$ \\
		\hline
	\end{tabular}
	
\end{table}
In Fig.\ref{fig:dsig_dpt} we show transverse momentum distribution
of one of the $\chi_{c}$ mesons for the $2 \to 2$
and $2 \to 3$ processes.
The $2 \to 3$ contributions were calculated with $p_{Tg} > 1 \, \rm{GeV}$. The factorization scale is chosen as $\mu_{f}= \sqrt{\hat{s}}$
and the energy in the center of mass of two protons is $8\: \rm{TeV}$.
We use the $MSTW2008nlo$ \cite{MSTW} parton distribution functions.
For illustration in a few plots we present only diagram (B)
from Fig.\ref{fig:diagrams_chicchicg}, since the behaviour of the process 
described by diagram (A) is exactly opposite. We discuss pairs, where two of
$\chi_c's$ have the same spin, it is good example to show all 
characteristics of these mesons. Notice, that in high transverse momentum region,
the $pp\rightarrow[\chi_c\chi_c]g$ process dominates for each $\chi_{c J}$, though for $\chi_{c1}$
it is not big effect.

\begin{figure}[!h]
\includegraphics[width=5.3cm]{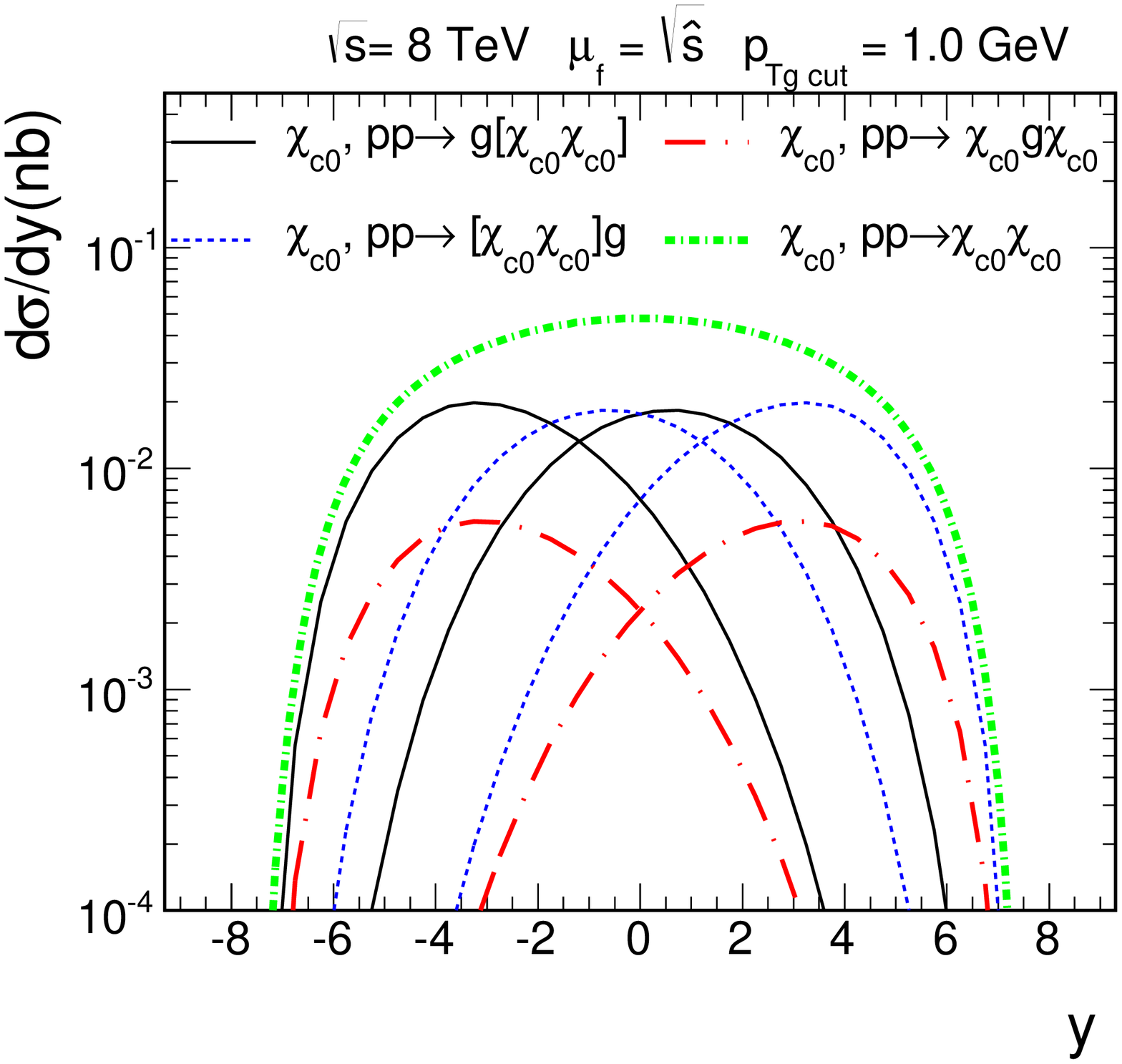}
\includegraphics[width=5.3cm]{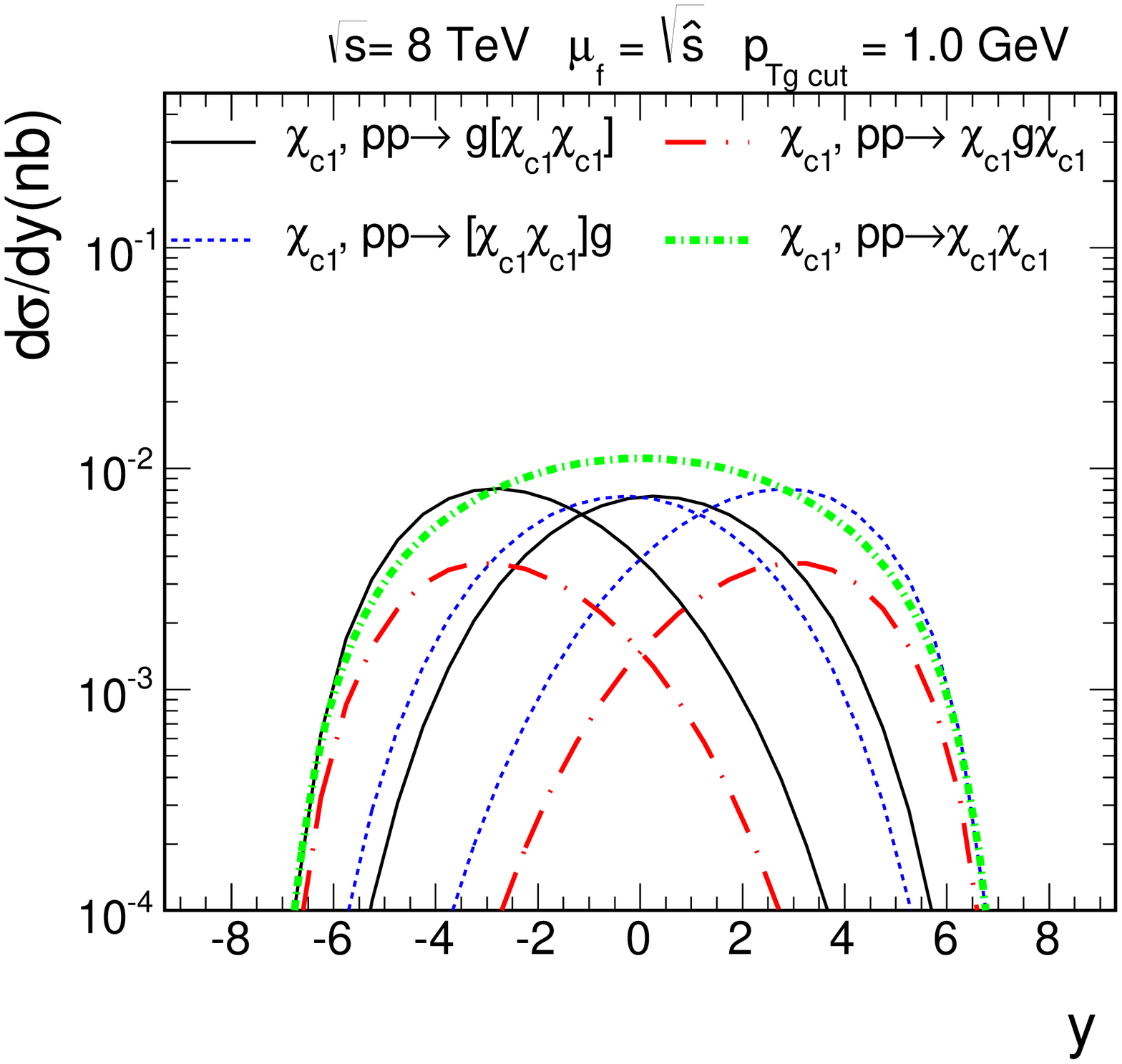}
\includegraphics[width=5.3cm]{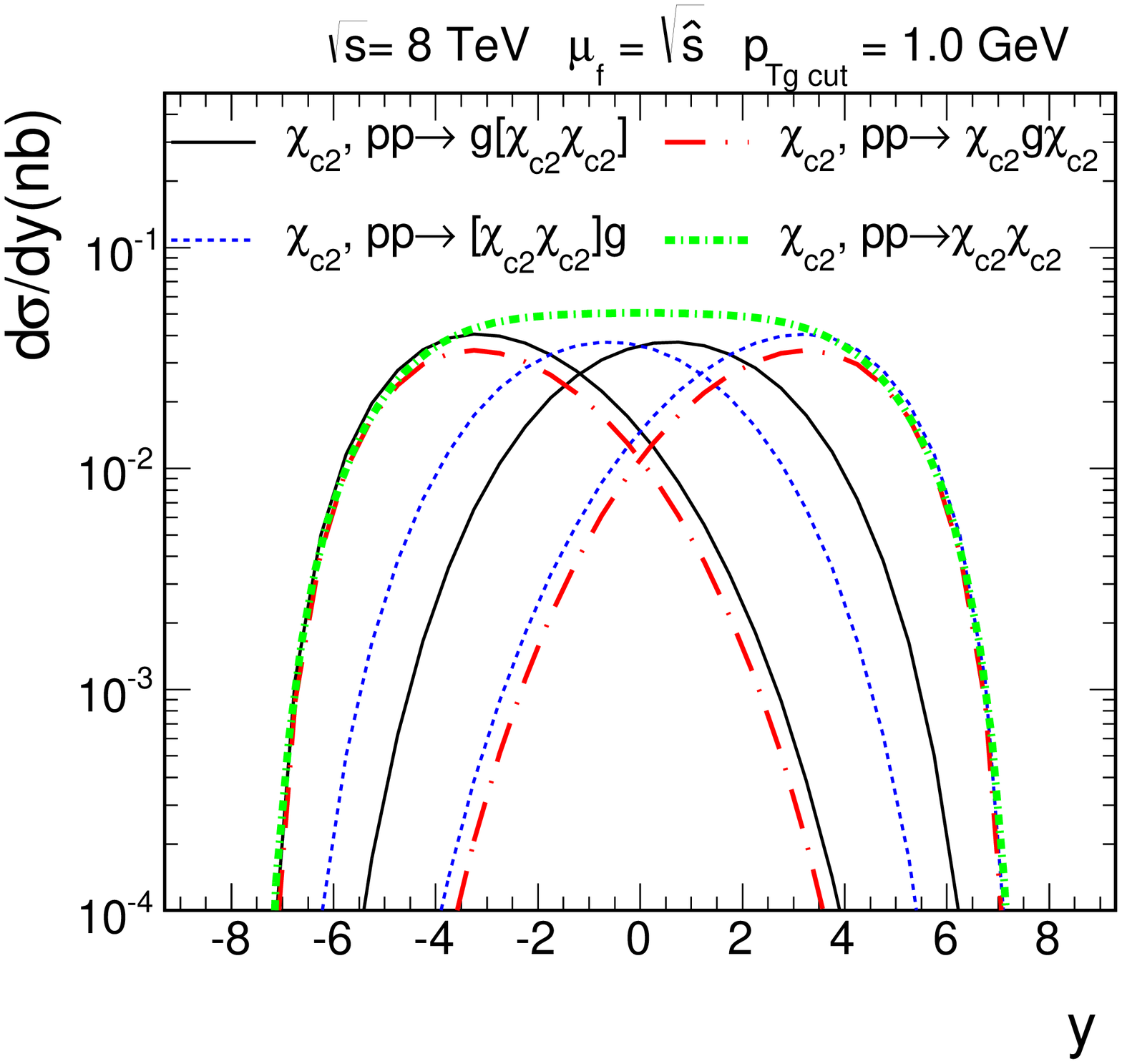}
\caption{Rapidity distributions of $\chi_c$ mesons from $2 \to 2$
and $2 \to 3$ processes shown in Fig.\ref{fig:diagrams_chicchicg}
for $\chi_{c 0} \chi_{c 0}$ (left) and $\chi_{c 1} \chi_{c 1}$ (middle)
and $\chi_{c 2} \chi_{c 2}$ (right).
Here $\mu_f^2 = \hat s$ was used.
}
\label{fig:ychic}
\end{figure}

In Fig.\ref{fig:ychic} we show rapidity distributions of $\chi_c$
mesons for the different $2 \to 2$ and $2 \to 3$ mechanisms discussed
above. One can see that in the rapidity range $|y|>3$
the $pp\rightarrow[\chi_{c1}\chi_{c1}]g$ process is not negligible
(the middle plot in Fig.\ref{fig:ychic}).
While the $2 \to 2$ sub-processes lead to the production of
$\chi_c$ mesons at midrapidities the $2 \to 3$ processes generate
$\chi_c$ mesons also at large $|y|$. Such mesons are then suppressed
in the midrapidity experiments as ATLAS or CMS. The same may be true
in the case of forward LHCb experiment. When the forward emitted meson 
is measured the second meson is emitted preferentially at midrapidities
(diagrams (A) and (B)) or even in opposite directions (diagram (C)).
We leave detailed studies relevant for a given experiment for the future.

\begin{figure}[!h]
	\includegraphics[width=5.3cm]{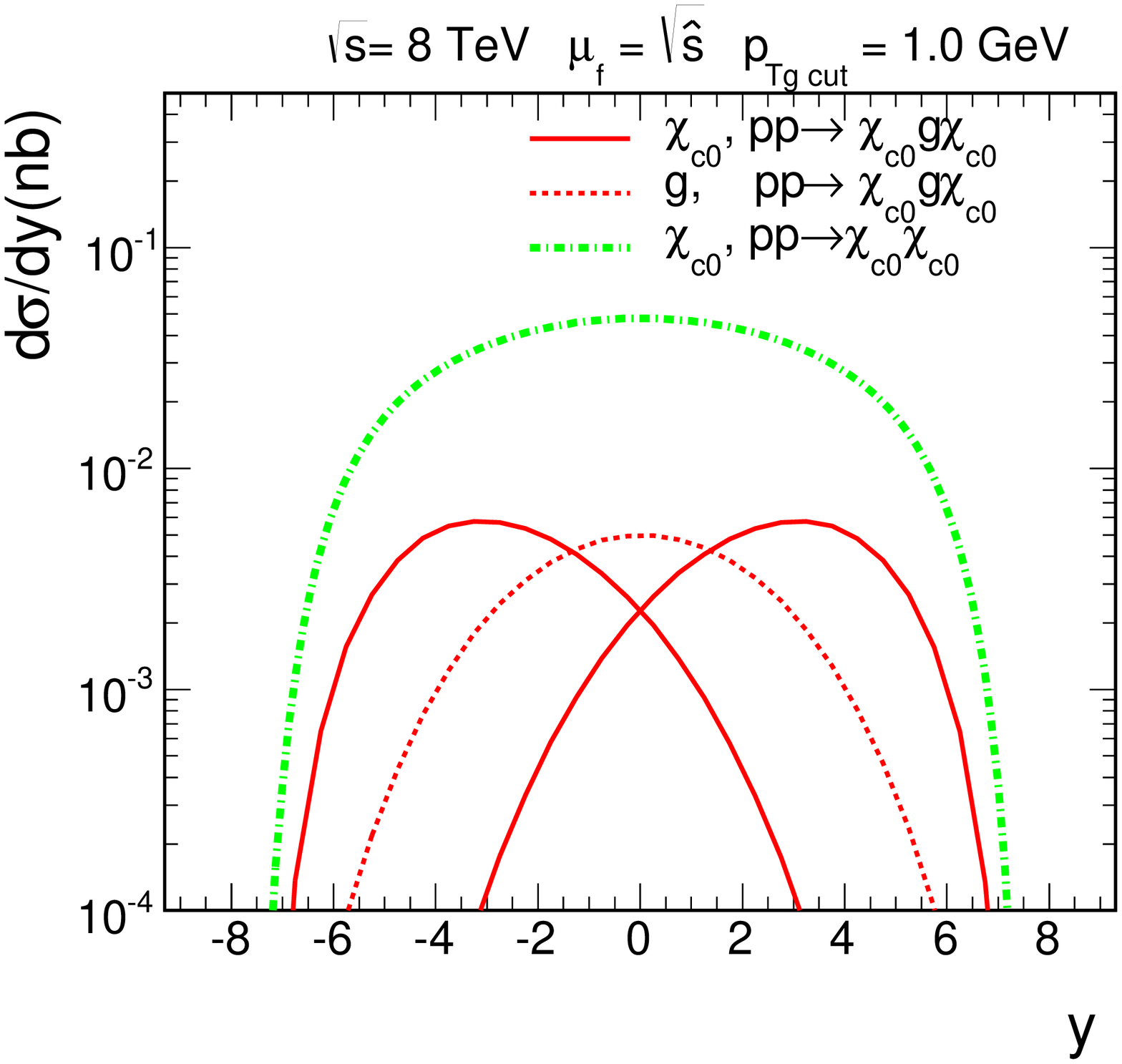}
	\includegraphics[width=5.3cm]{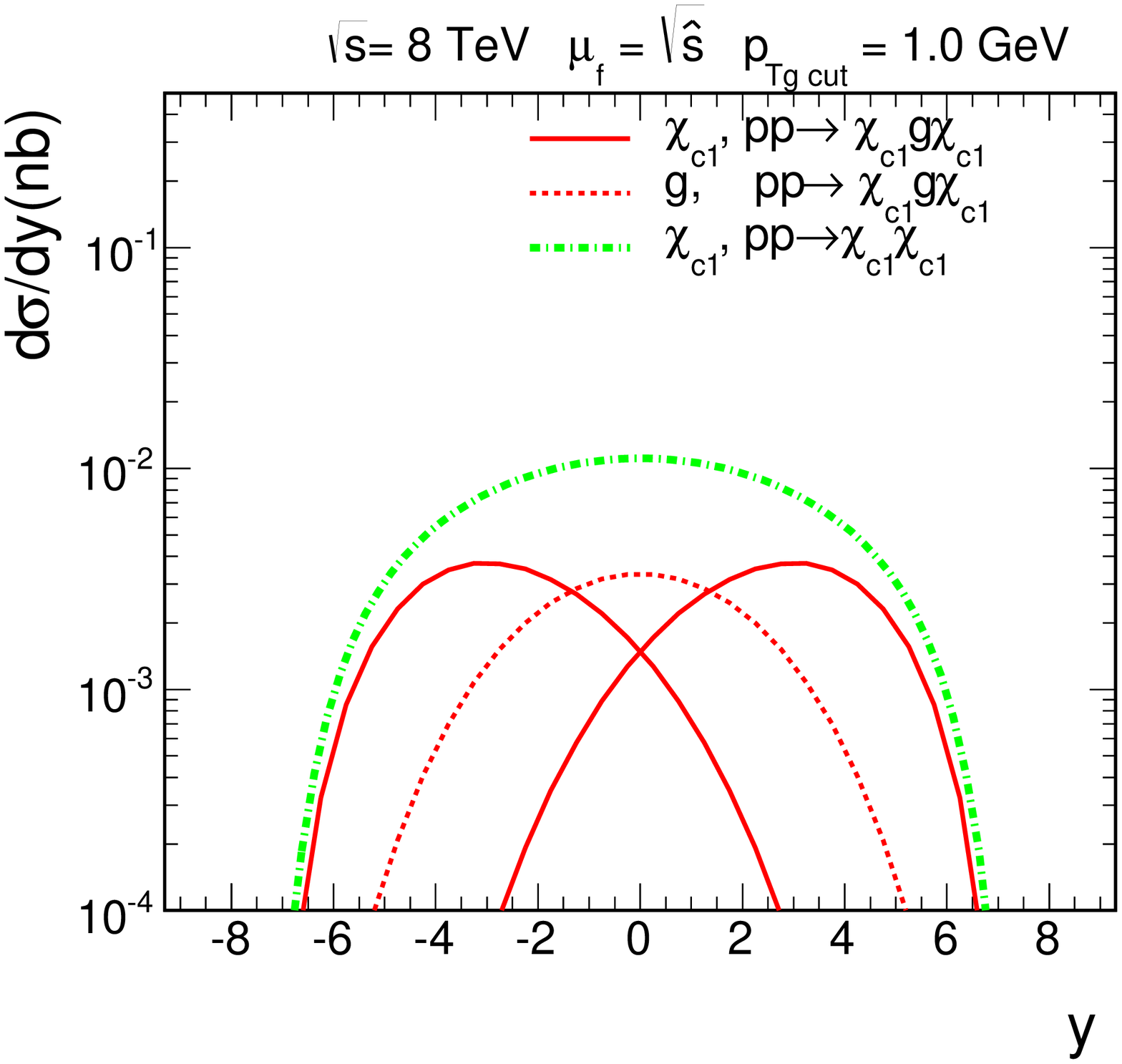}
	\includegraphics[width=5.3cm]{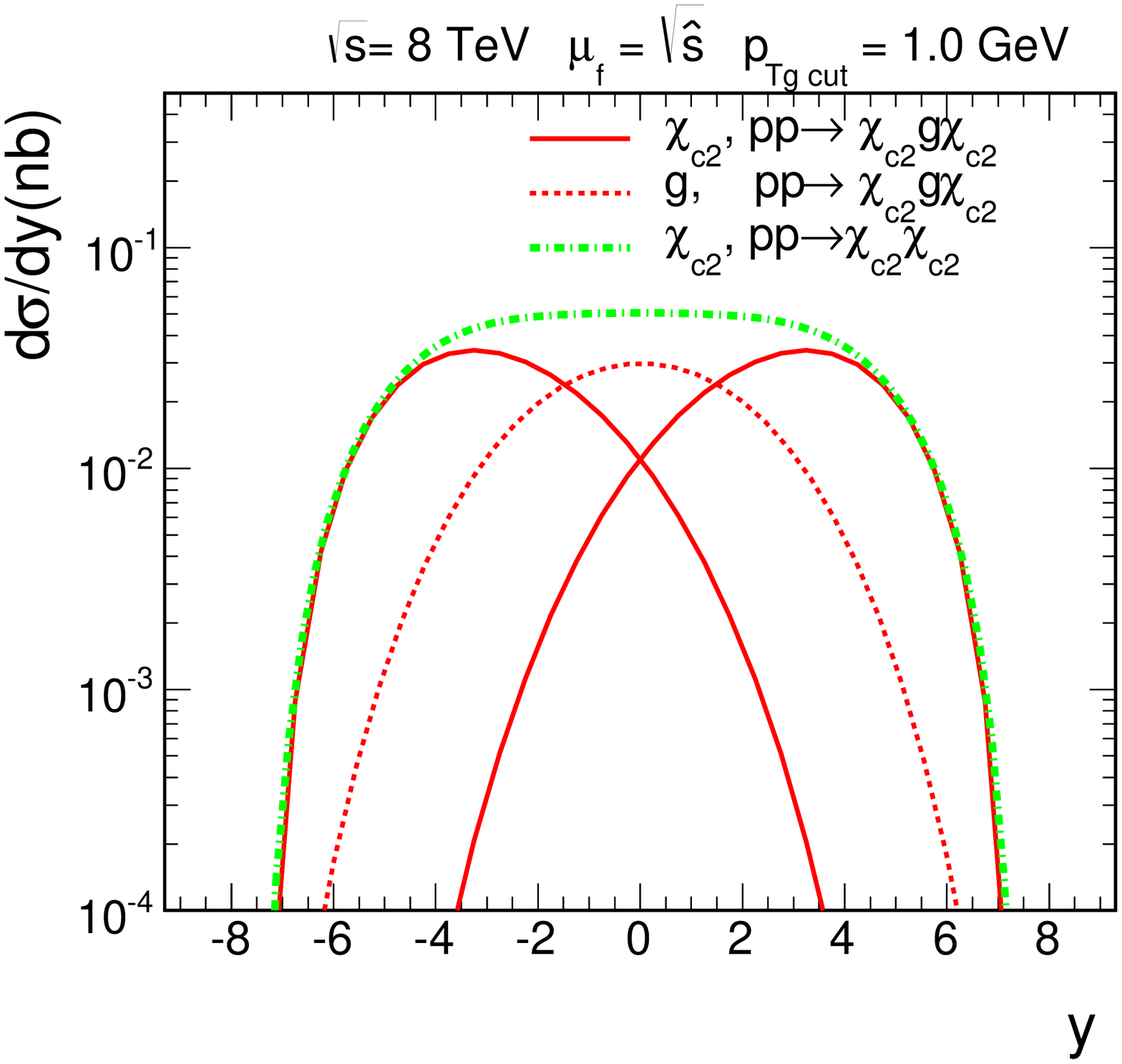}
	\caption{Rapidity distributions of $\chi_c$ mesons and gluons from the $ gg \to \chi g \chi$ processes with a central gluon. Here $\mu_f^2 = \hat s$ was used.
	}
	\label{fig:ychicandg_diagramc}
\end{figure}

In Fig.\ref{fig:ychicandg_diagramc} we compare rapidity distributions
of $\chi_c$ mesons and the associated gluon (see diagram (C) in
Fig.\ref{fig:diagrams_chicchicg}). In this case, while the $\chi_c$ quarkonia are produced 
preferentially in forward or backward directions, gluons are emitted 
preferentially at midrapidities. For comparison we show also
distributions of $\chi_c$ quarkonia from the $2 \to 2$ sub-processes.

\begin{figure}[!h]
	\includegraphics[width=5.3cm]{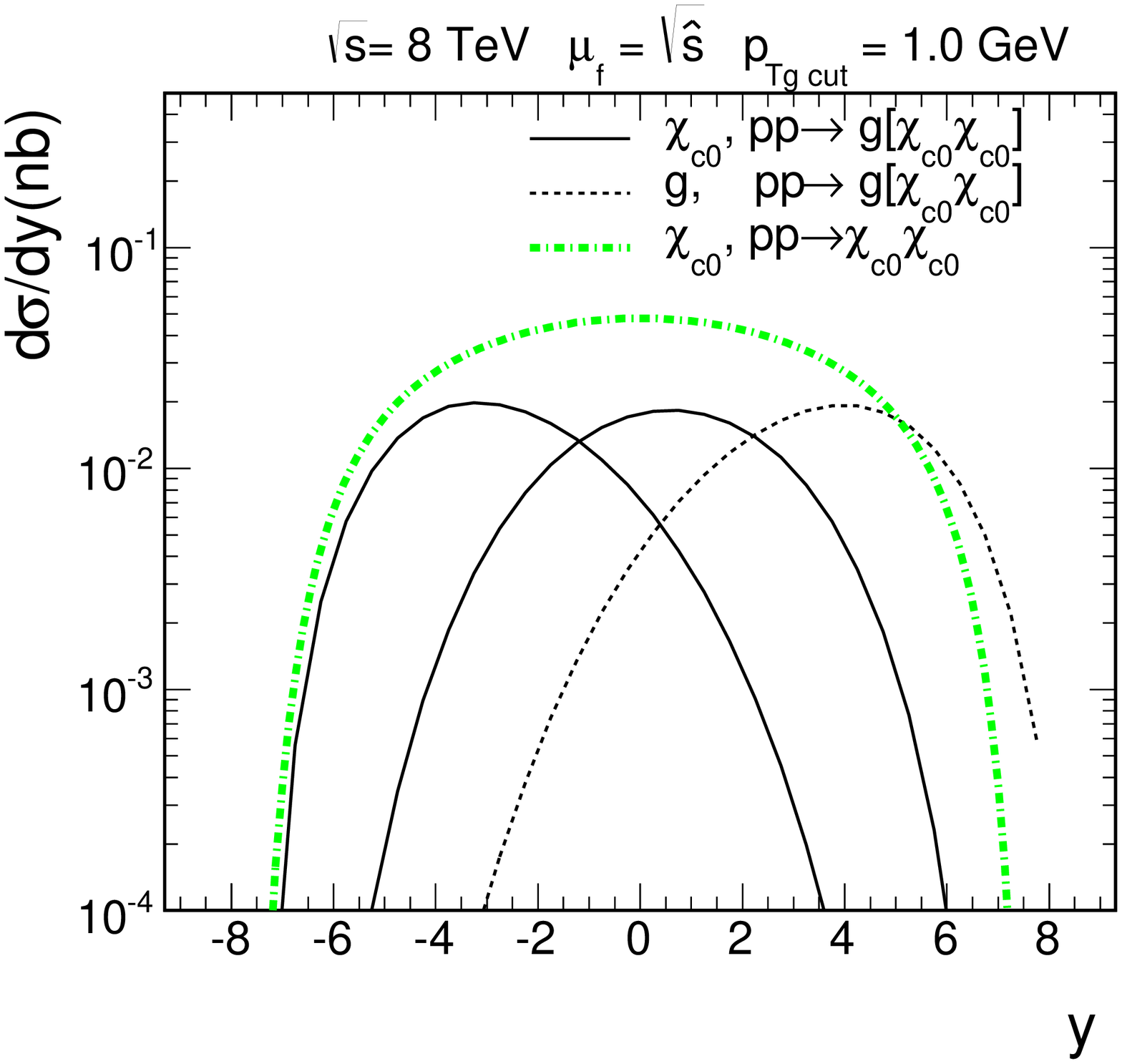}
	\includegraphics[width=5.3cm]{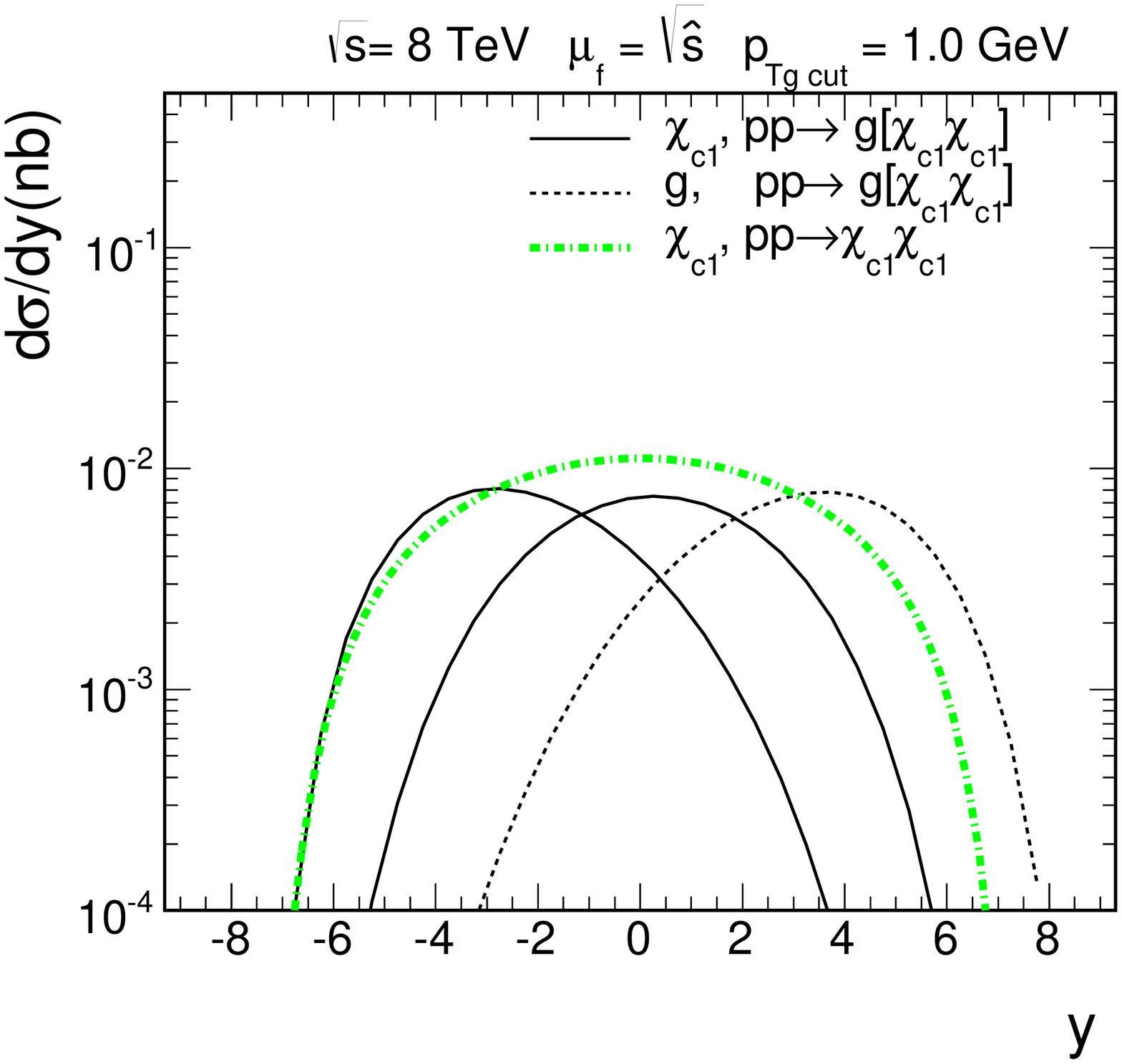}
	\includegraphics[width=5.3cm]{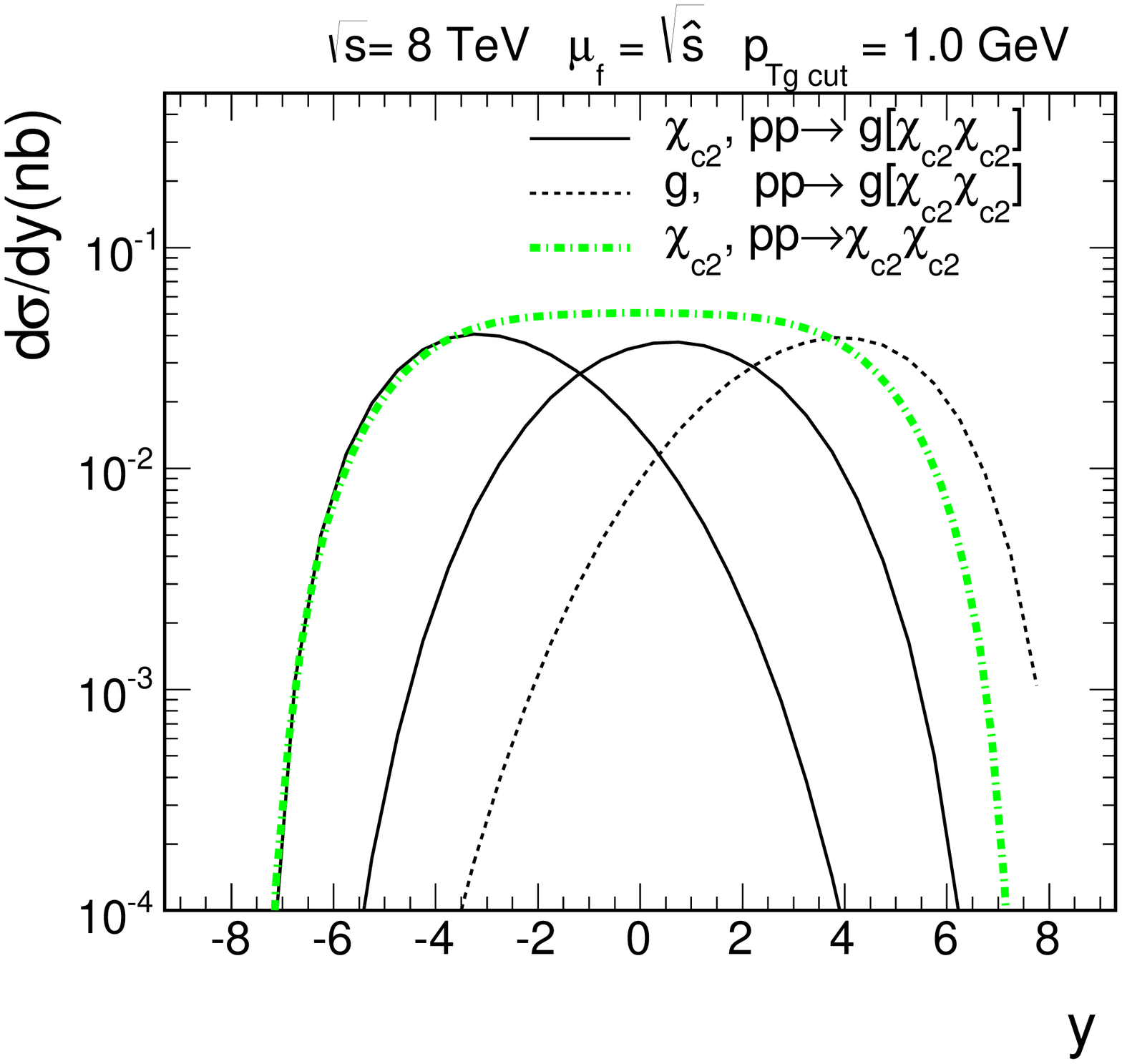}
	\caption{Rapidity distributions of $\chi_c$ mesons and gluons from the $ gg \to g \chi_c  \chi_c$ processes with a leading gluon. Here $\mu_f^2 = \hat s$ was used.
	}
	\label{fig:ychicandg_diagrama}
\end{figure}

In Fig.\ref{fig:ychicandg_diagrama} we show similar distributions
for diagrams (A) and (B) in Fig.\ref{fig:diagrams_chicchicg} and for
reference also the distributions from the $2 \to 2$ sub-processes.

In general, there is rapidity ordering of final state particles
for the considered $2 \to 3$ processes.
To see it even better let us present now distributions in rapidity 
differences between final state objects.

\begin{figure}[!h]
\includegraphics[width=8cm]{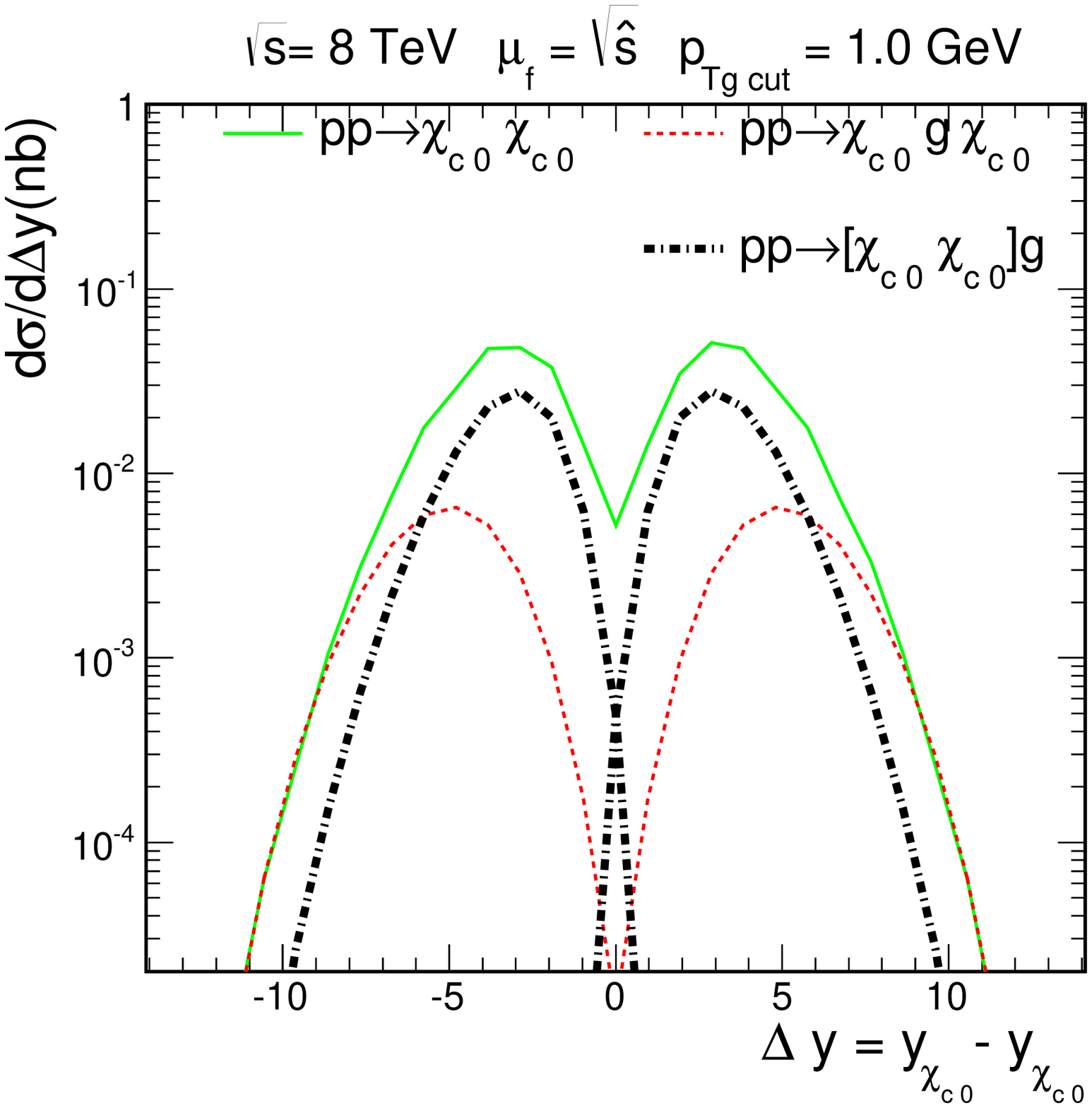}
\includegraphics[width=8cm]{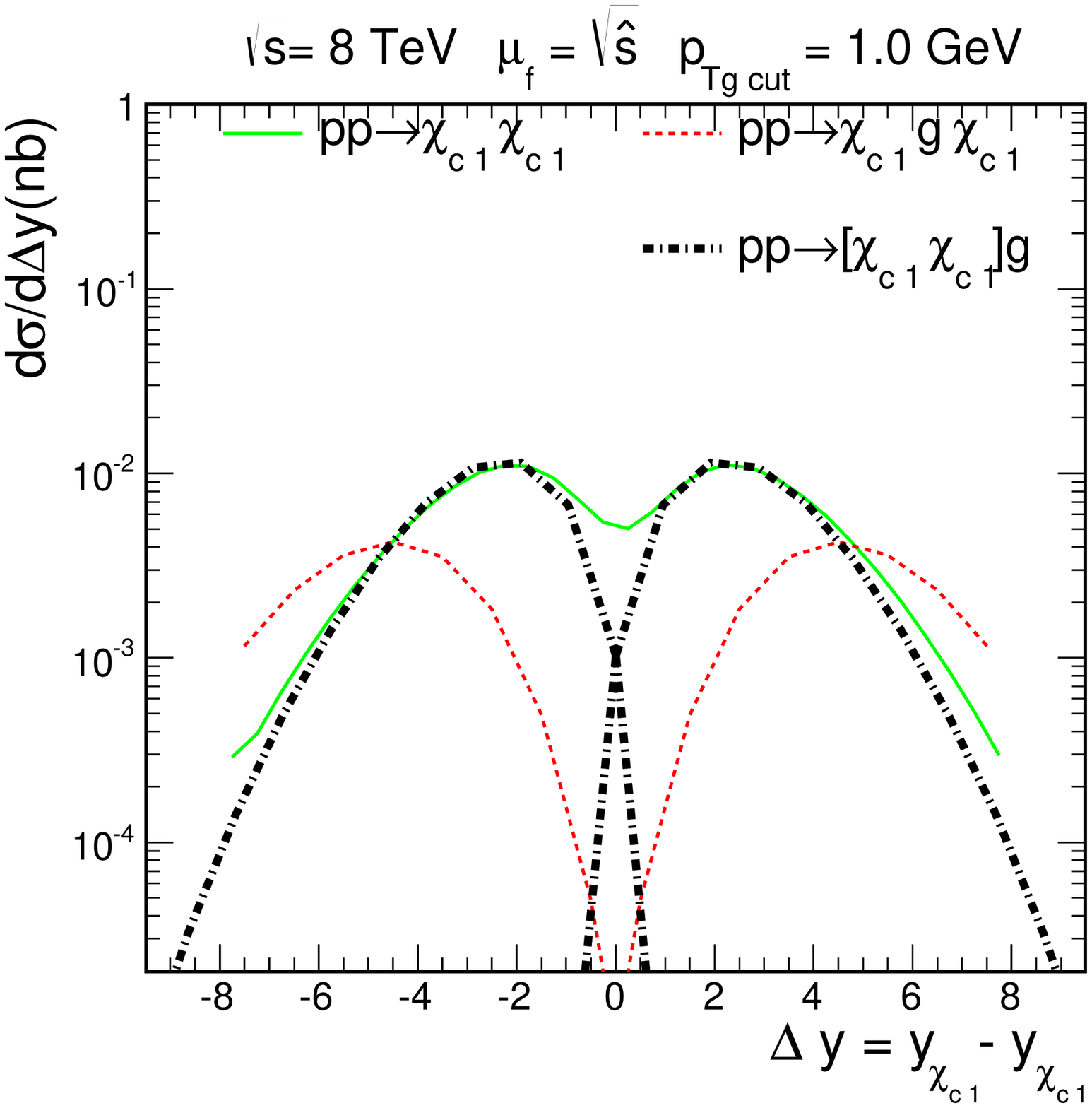}
\includegraphics[width=8cm]{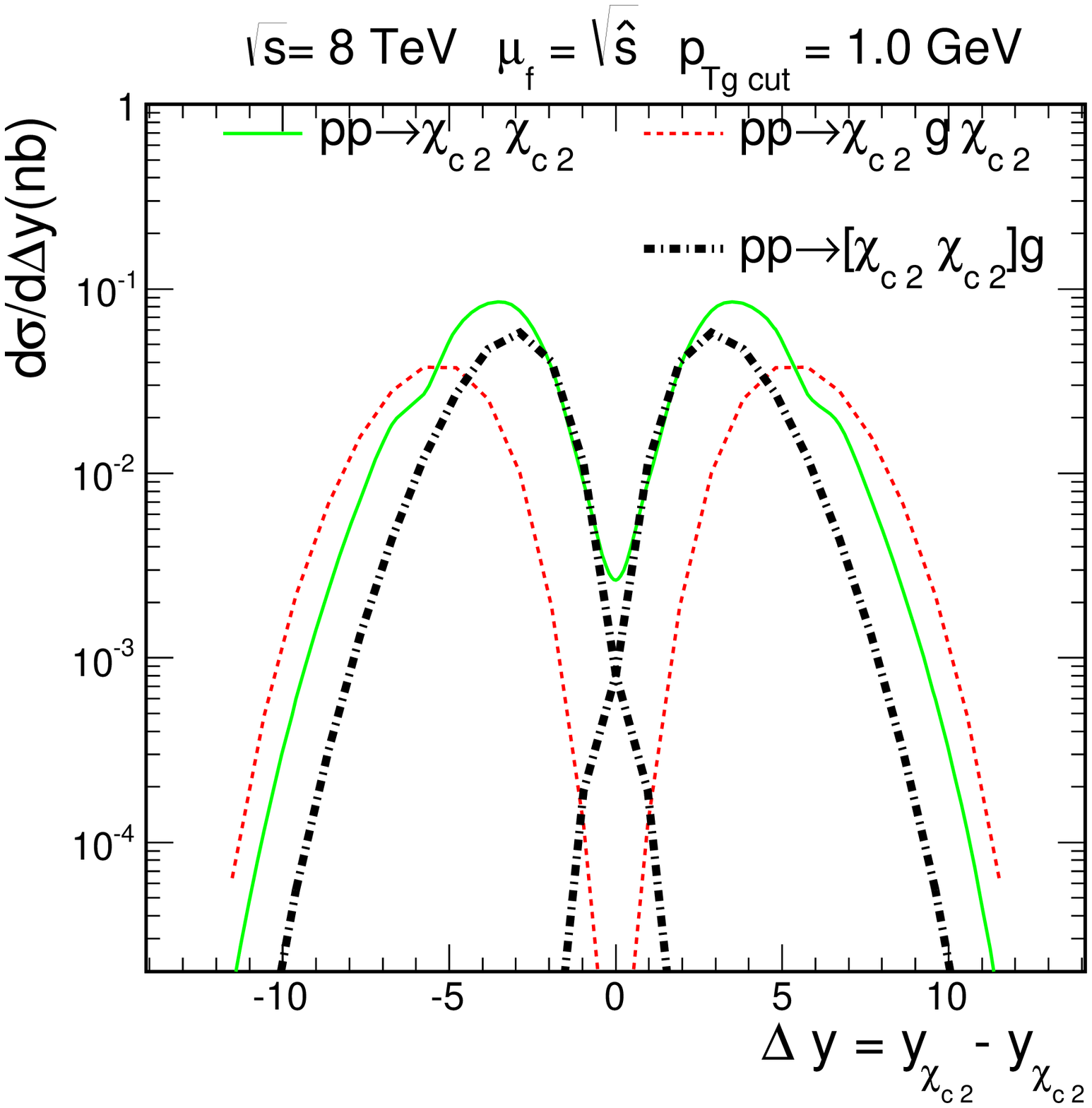}
\caption{Distribution in the difference of rapidities of $\chi_{c J}$
mesons from processes shown in diagrams [\ref{fig:diagrams_chicchicg}].
The most external lines are from process, where gluon is emitted among
the two $\chi_{c J}$ mesons.}
\label{fig:dsig_dydiffchic_chic}
\end{figure}

The distribution in rapidity distance between two $\chi_c$ mesons
is shown in Fig.\ref{fig:dsig_dydiffchic_chic} for different components discussed
in the present paper: $\chi_{c 0} \chi_{c 0}$, $\chi_{c 1} \chi_{c 1}$ and 
$\chi_{c 2} \chi_{c 2}$.
Indeed, as expected, the largest distances between the $\chi_c$
quarkonia are populated by processes with the gluon emitted among 
both $\chi_c$ mesons. 
Then also a sizeable gap at small rapidity distances can be
observed.

\begin{figure}
\includegraphics[width=8cm]{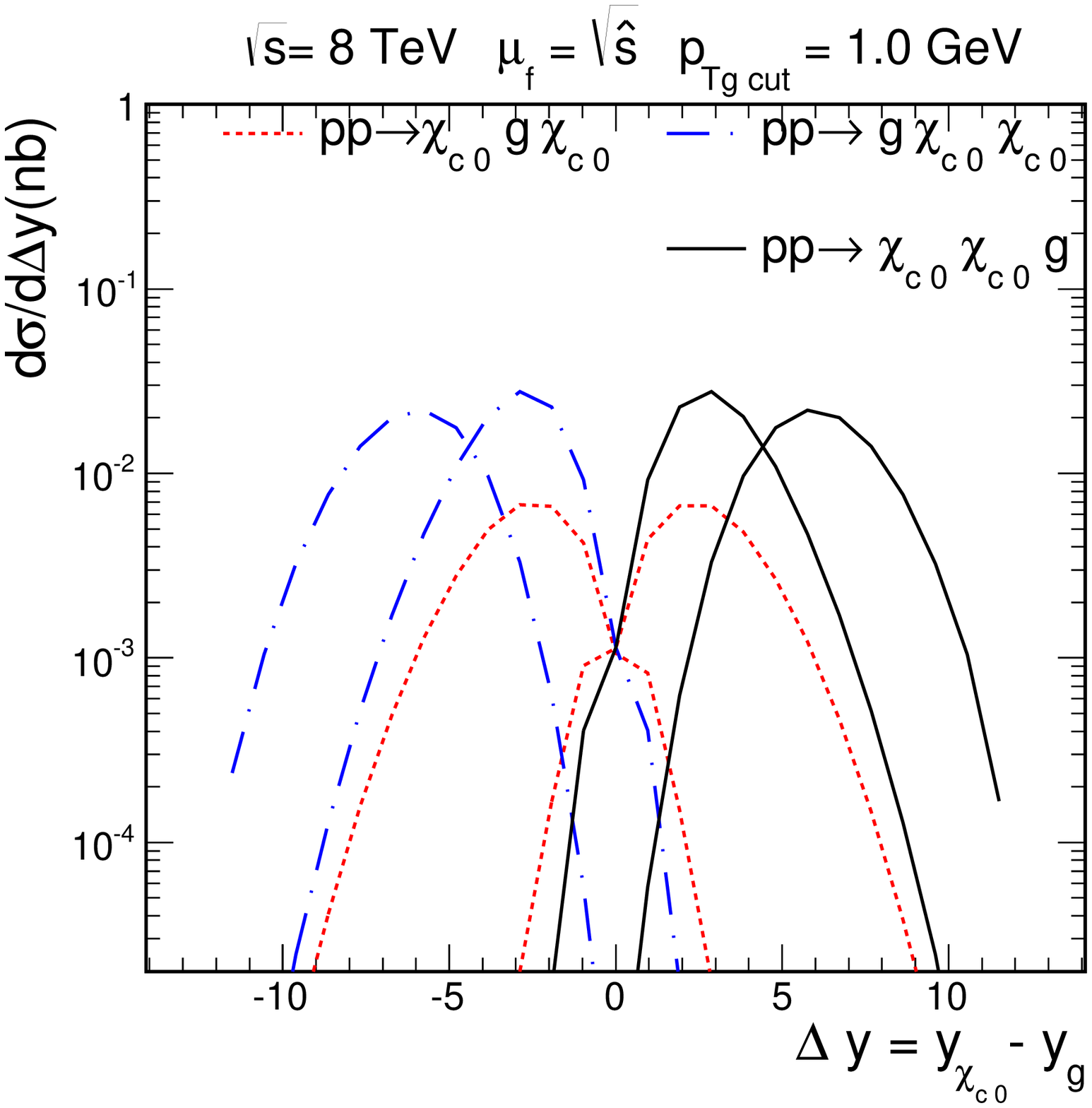}
\includegraphics[width=8cm]{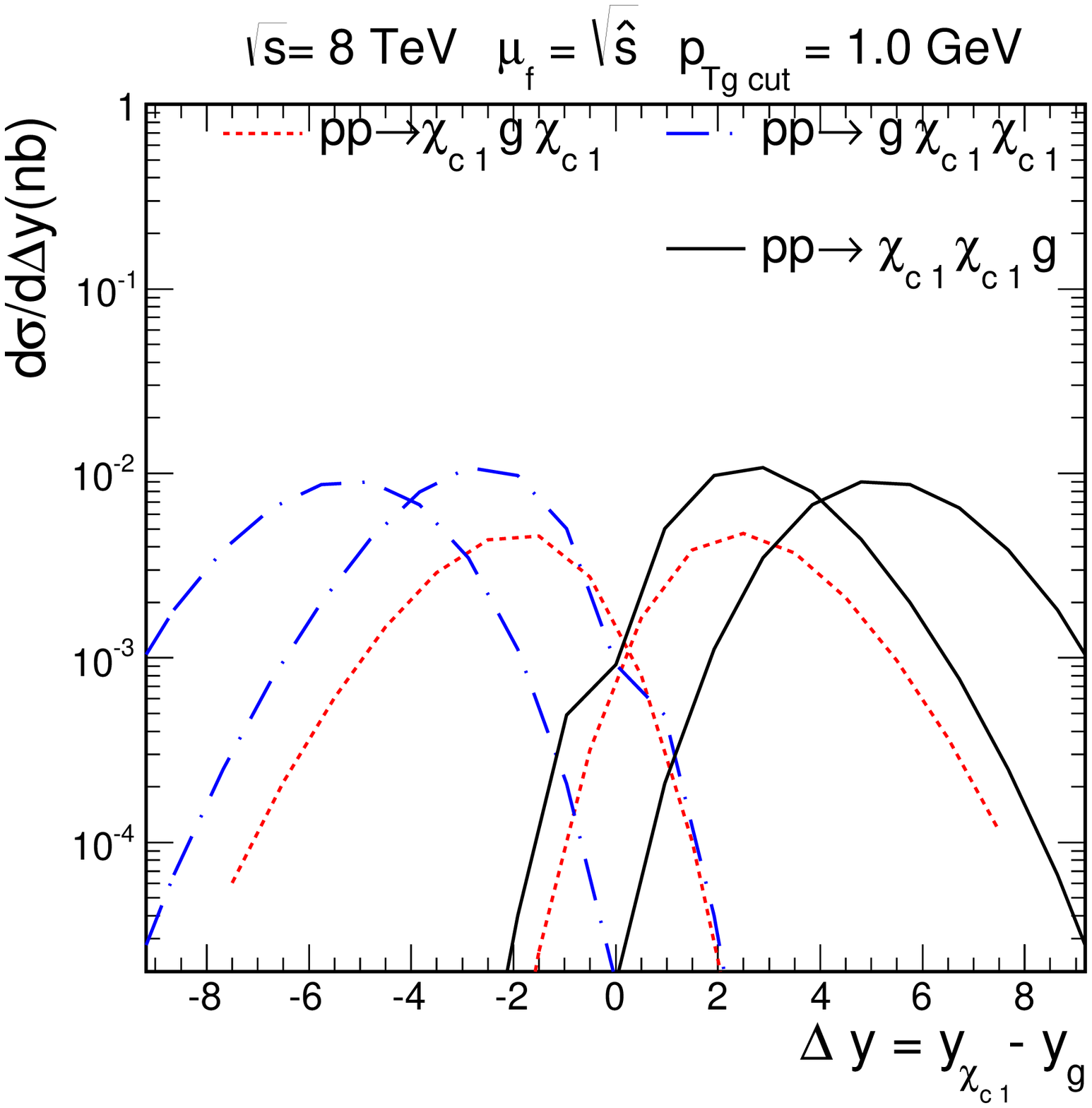}
\includegraphics[width=8cm]{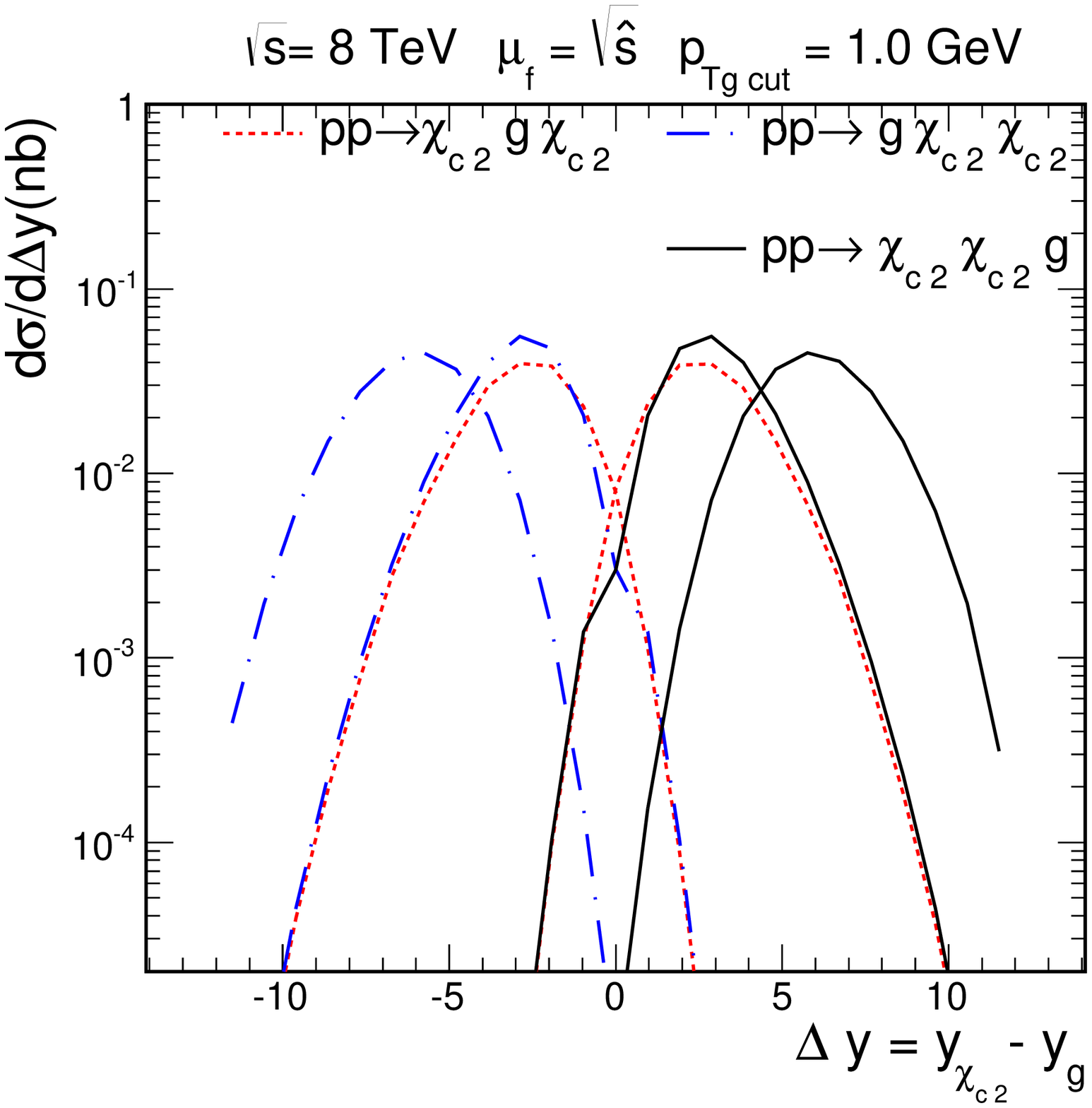}
\caption{Distribution in the difference of rapidities between $\chi_{c J}$
meson and gluon from diagrams presented in Fig.\ref{fig:diagrams_chicchicg}.
The most external lines in the plot are for difference between external 
gluon and the external $\chi_{c J}$ meson
(diagram (A) and (B) from Fig. \ref{fig:diagrams_chicchicg} ).  }
\label{fig:dsig_dydiffchic_g}
\end{figure}

In Fig.\ref{fig:dsig_dydiffchic_g} we show similar distributions, this time for
rapidity distance between one of the $\chi_c$ mesons and the associated 
gluon for $\chi_{c 0} \chi_{c 0}$, $\chi_{c 1} \chi_{c 1}$ and 
$\chi_{c 2} \chi_{c 2}$.
The considered mechanisms prefer large distances also in this variable.


Let us discuss now some correlation observables.

\begin{figure}[!h]
\includegraphics[width=5.3cm]{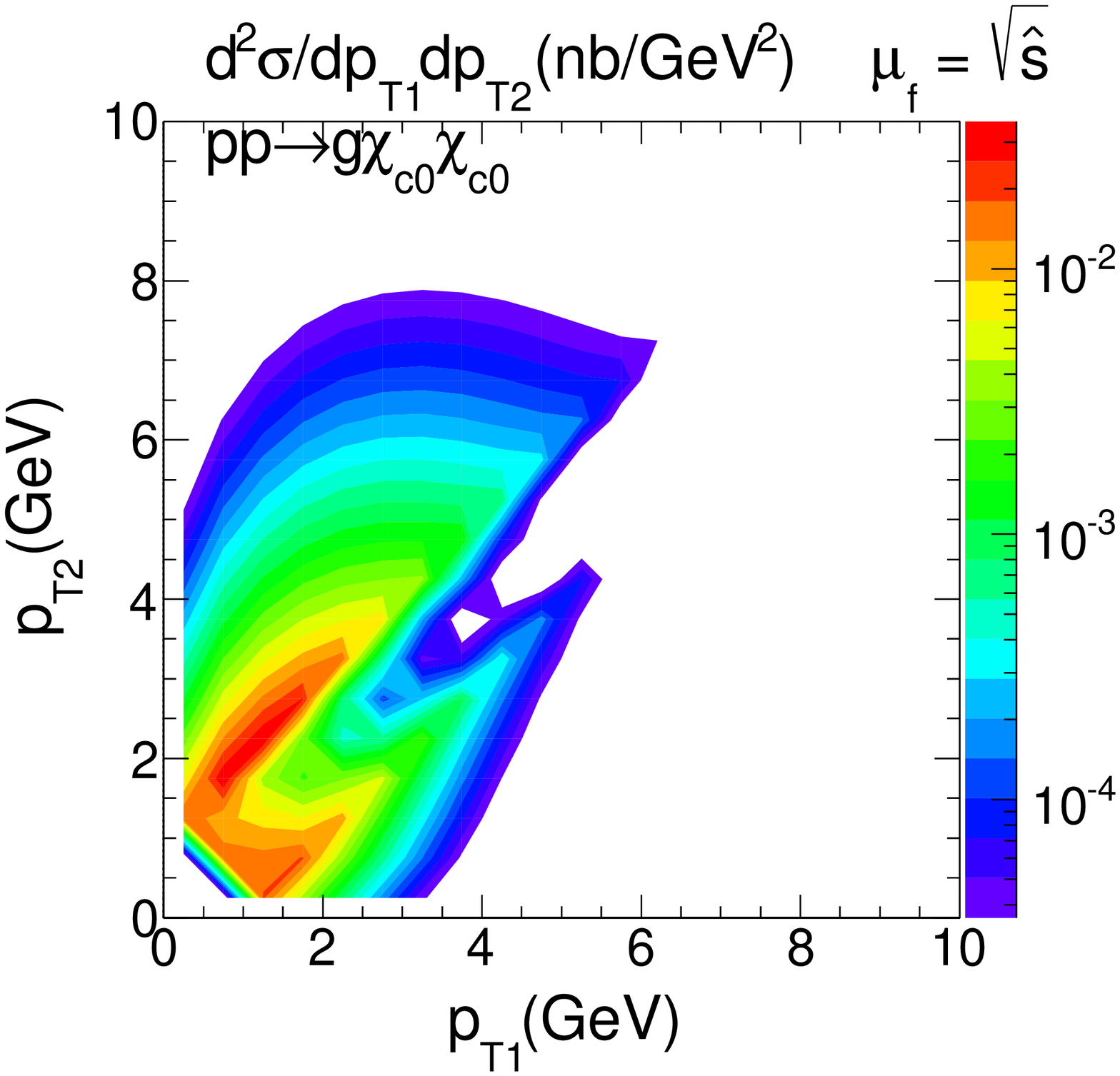}
\includegraphics[width=5.3cm]{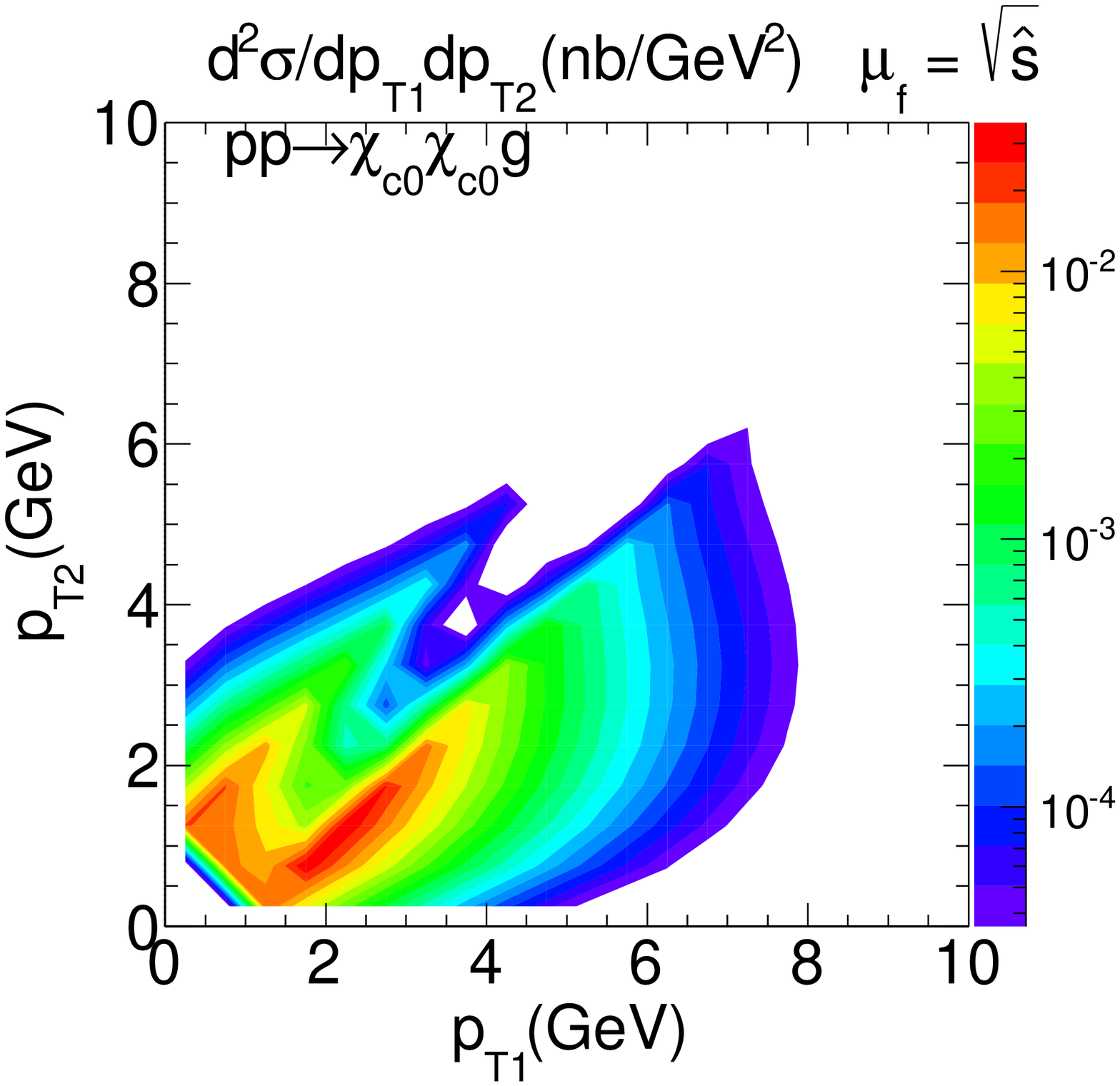}
\includegraphics[width=5.3cm]{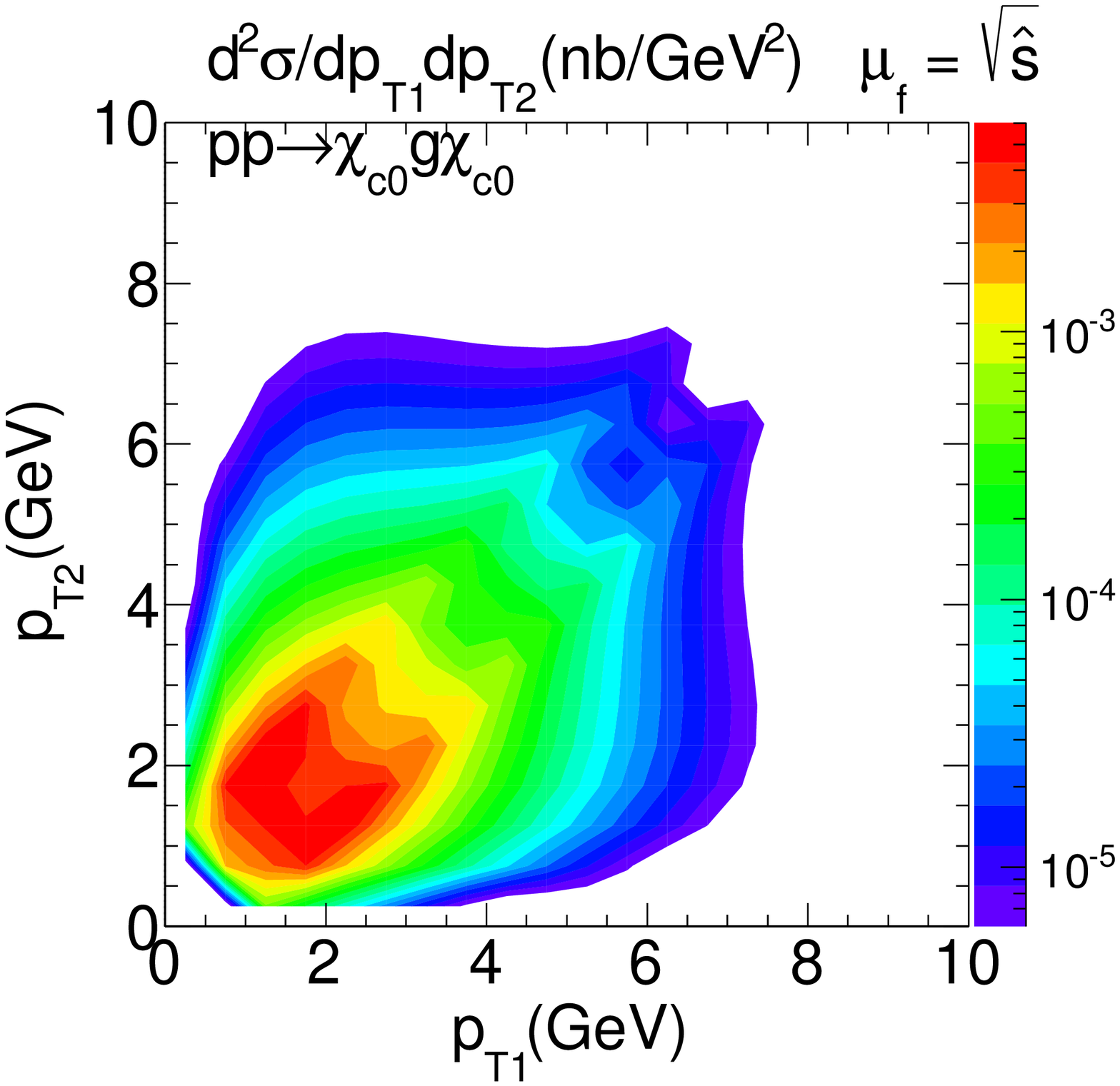}
\caption{Two-dimensional distributions ($p_{T1} \times p_{T2}$) for $\chi_{c 0} \chi_{c 0}$ production. Here $\mu_f^2 = \hat s$.}

\label{fig:p1tp2t_chic0chic0}
\end{figure}

\begin{figure}[!h]
\includegraphics[width=5.3cm]{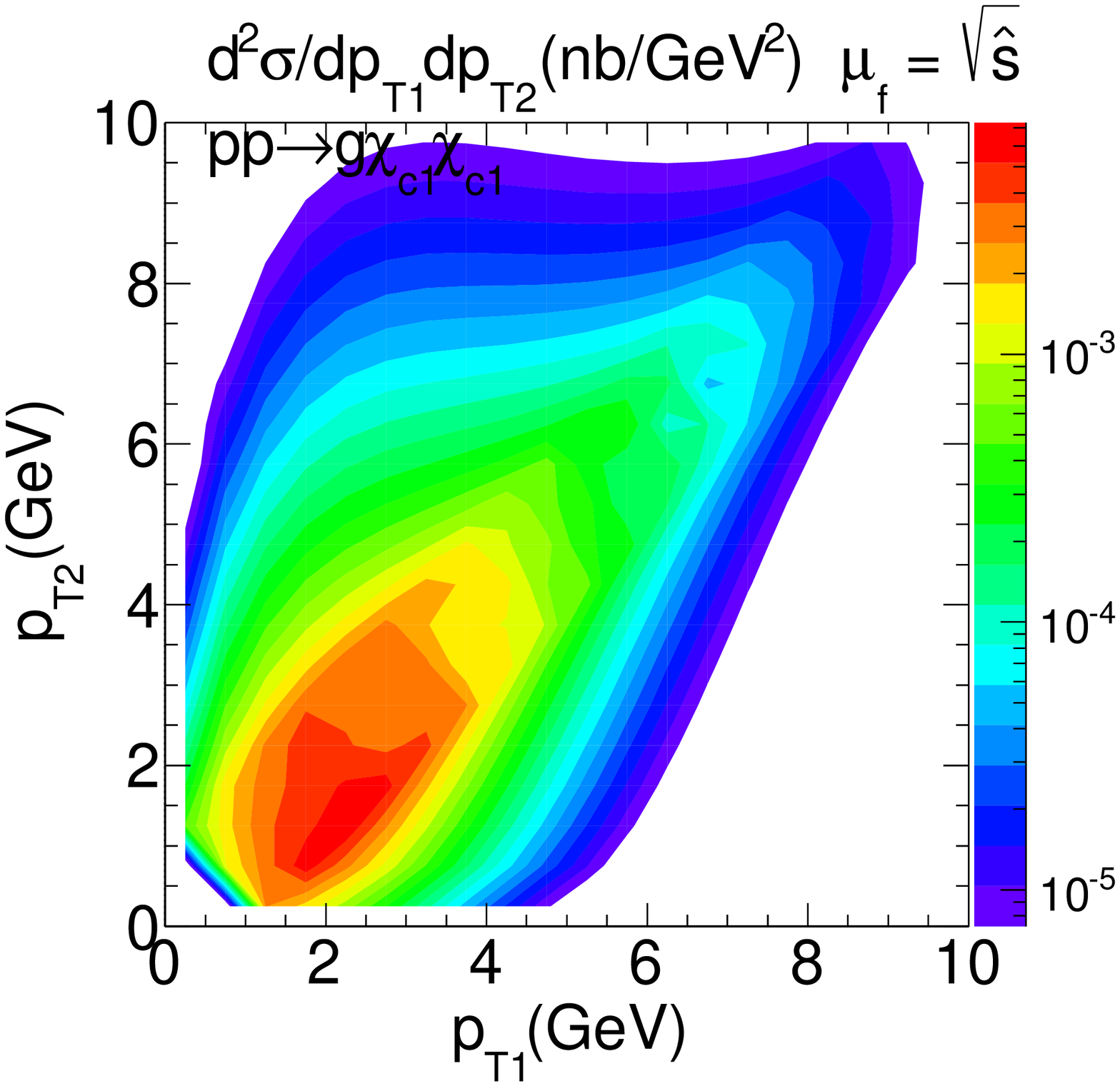}
\includegraphics[width=5.3cm]{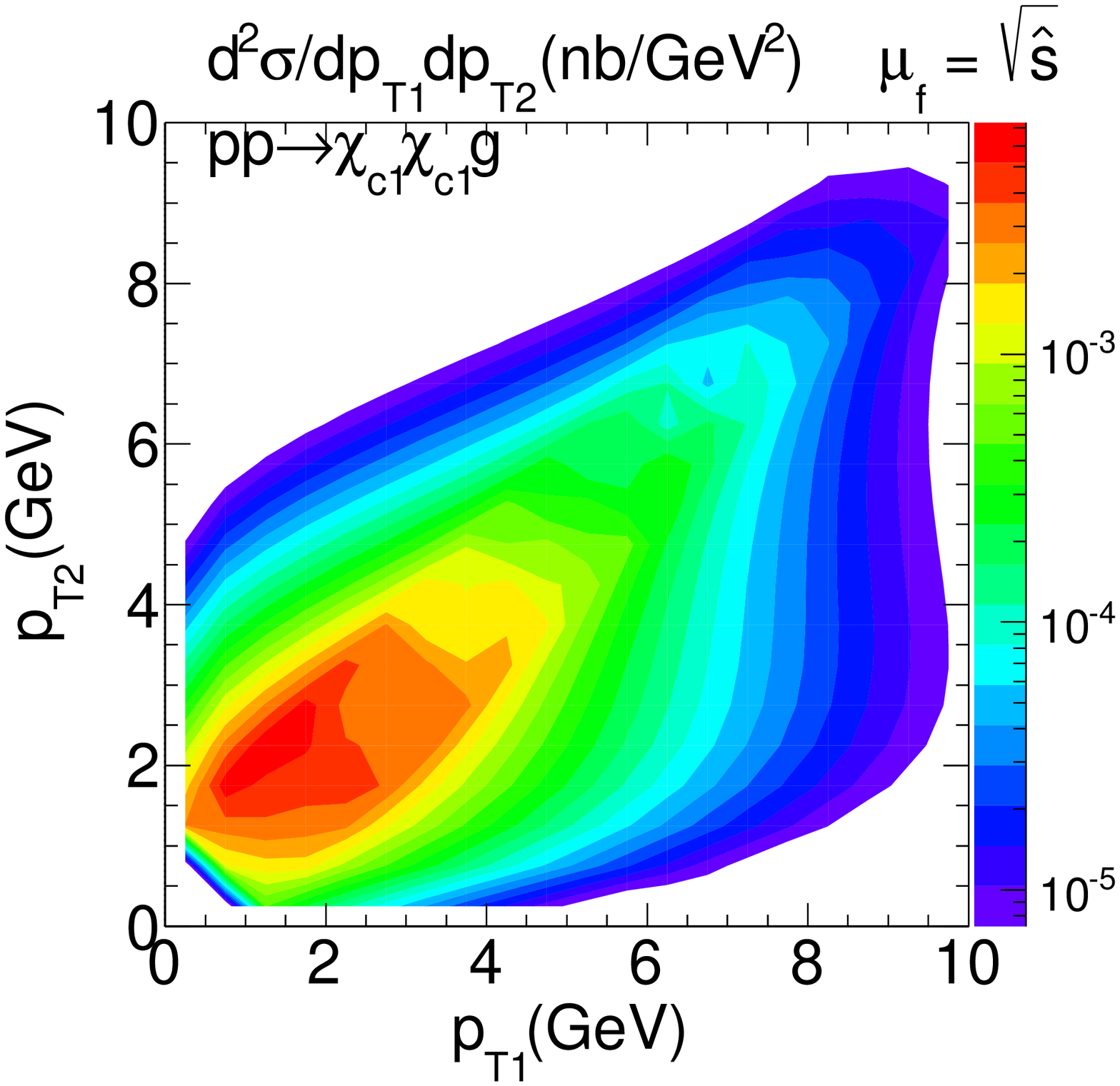}
\includegraphics[width=5.3cm]{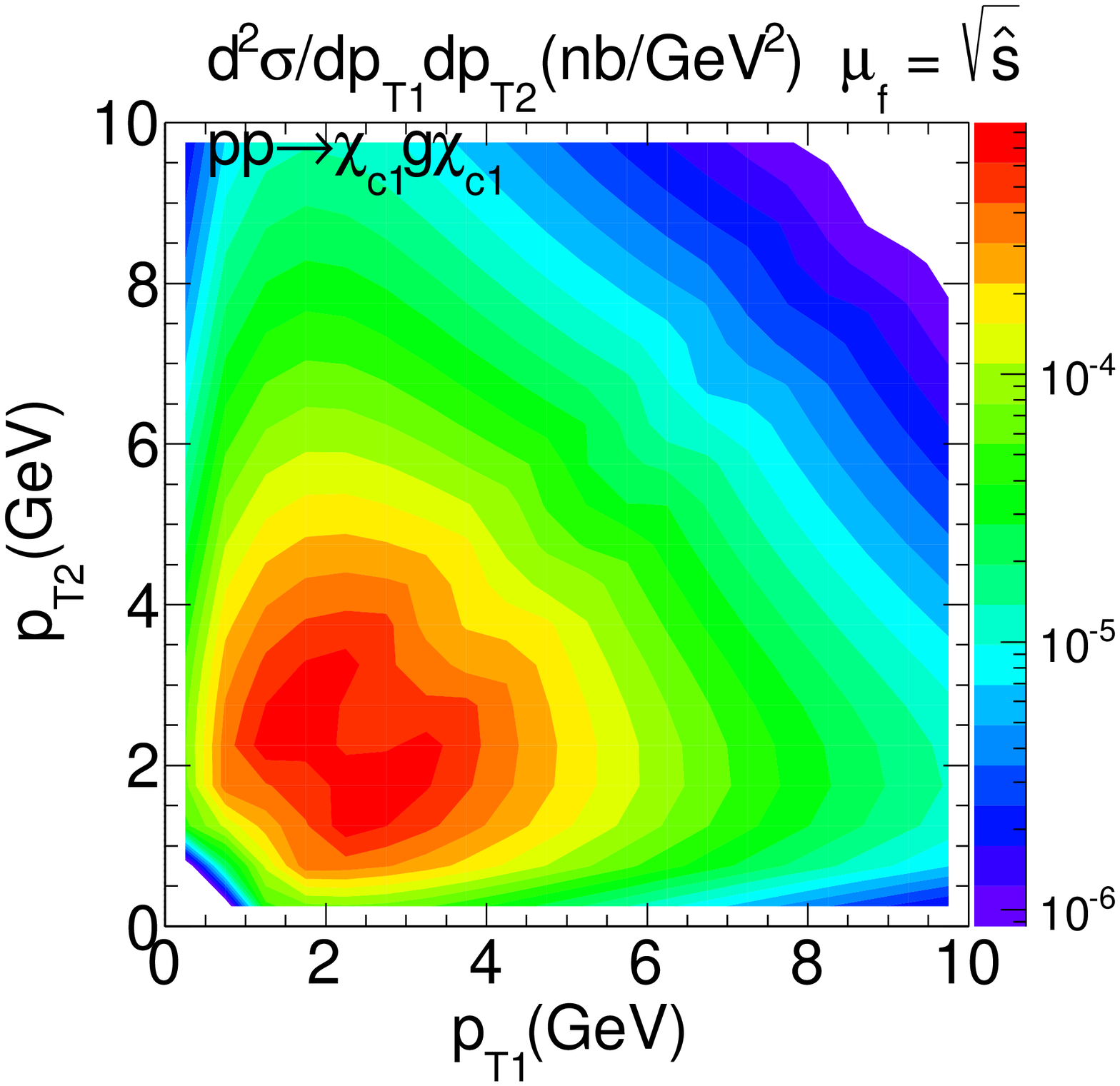}
\caption{Two-dimensional distributions ($p_{T1} \times p_{T2}$)
for $\chi_{c 1} \chi_{c 1}$ production.
Here $\mu_f^2 = \hat s$.
}
\label{fig:p1tp2t_chic1chic1}
\end{figure}

\begin{figure}[!h]
\includegraphics[width=5.3cm]{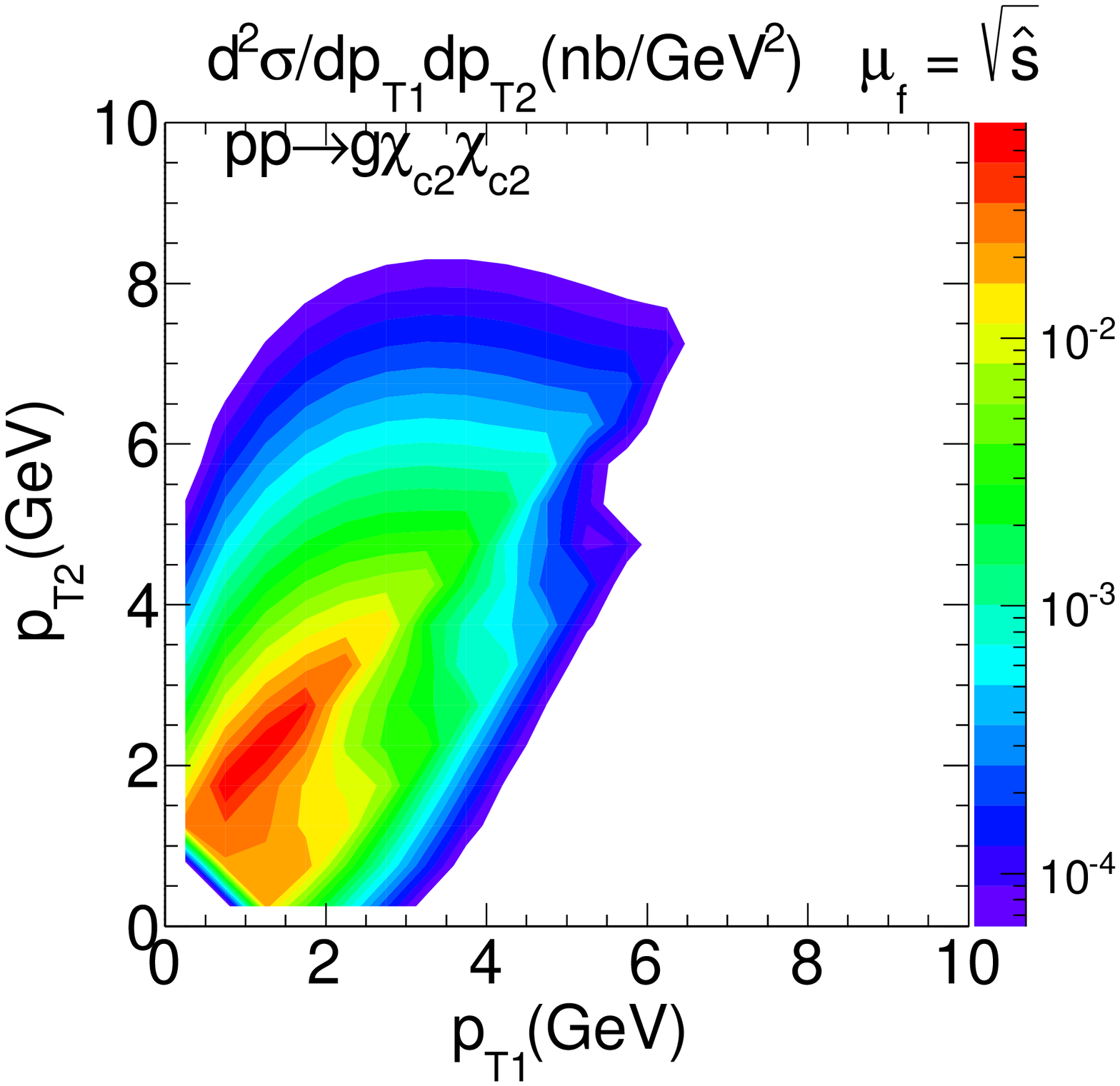}
\includegraphics[width=5.3cm]{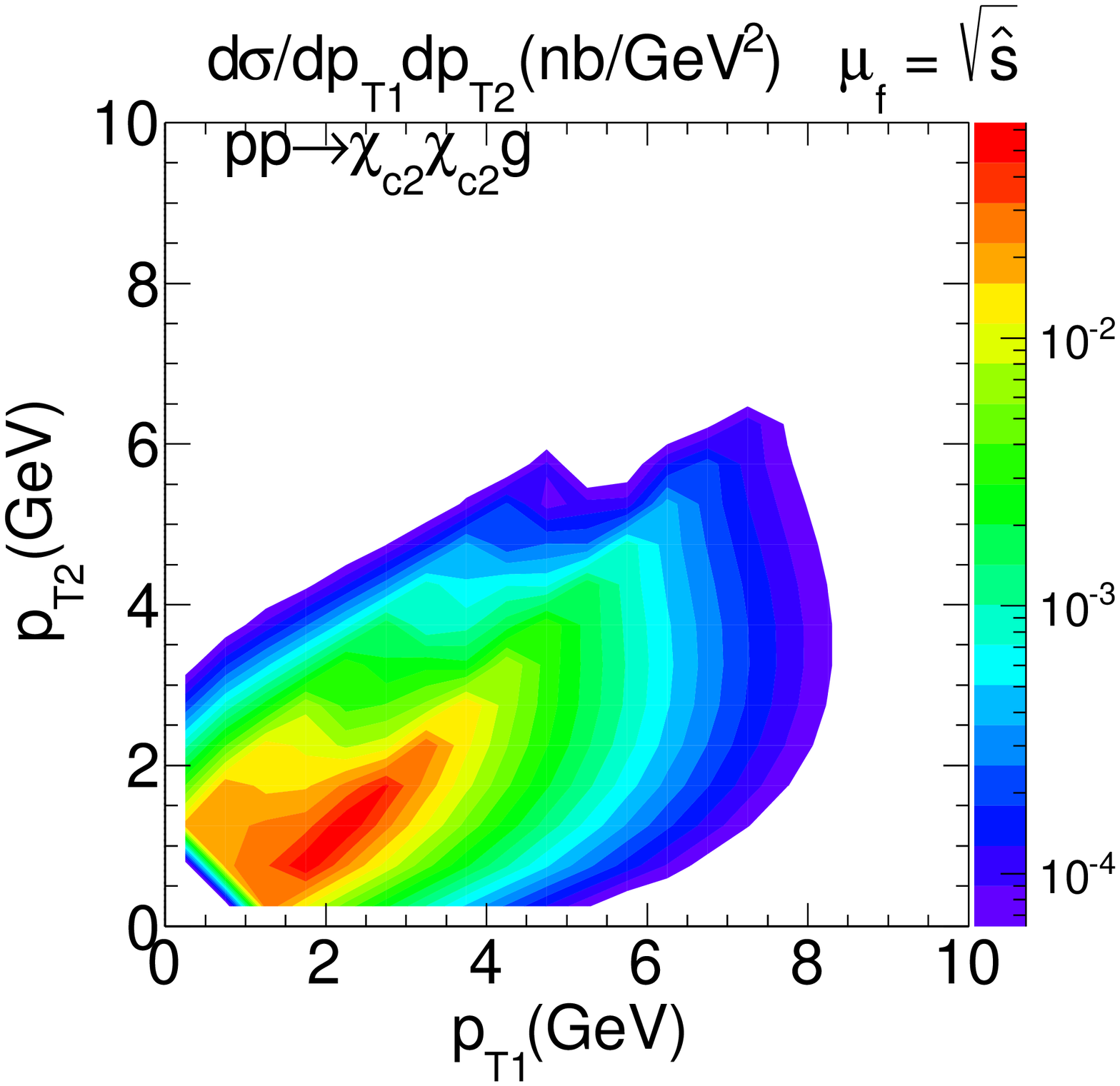}
\includegraphics[width=5.3cm]{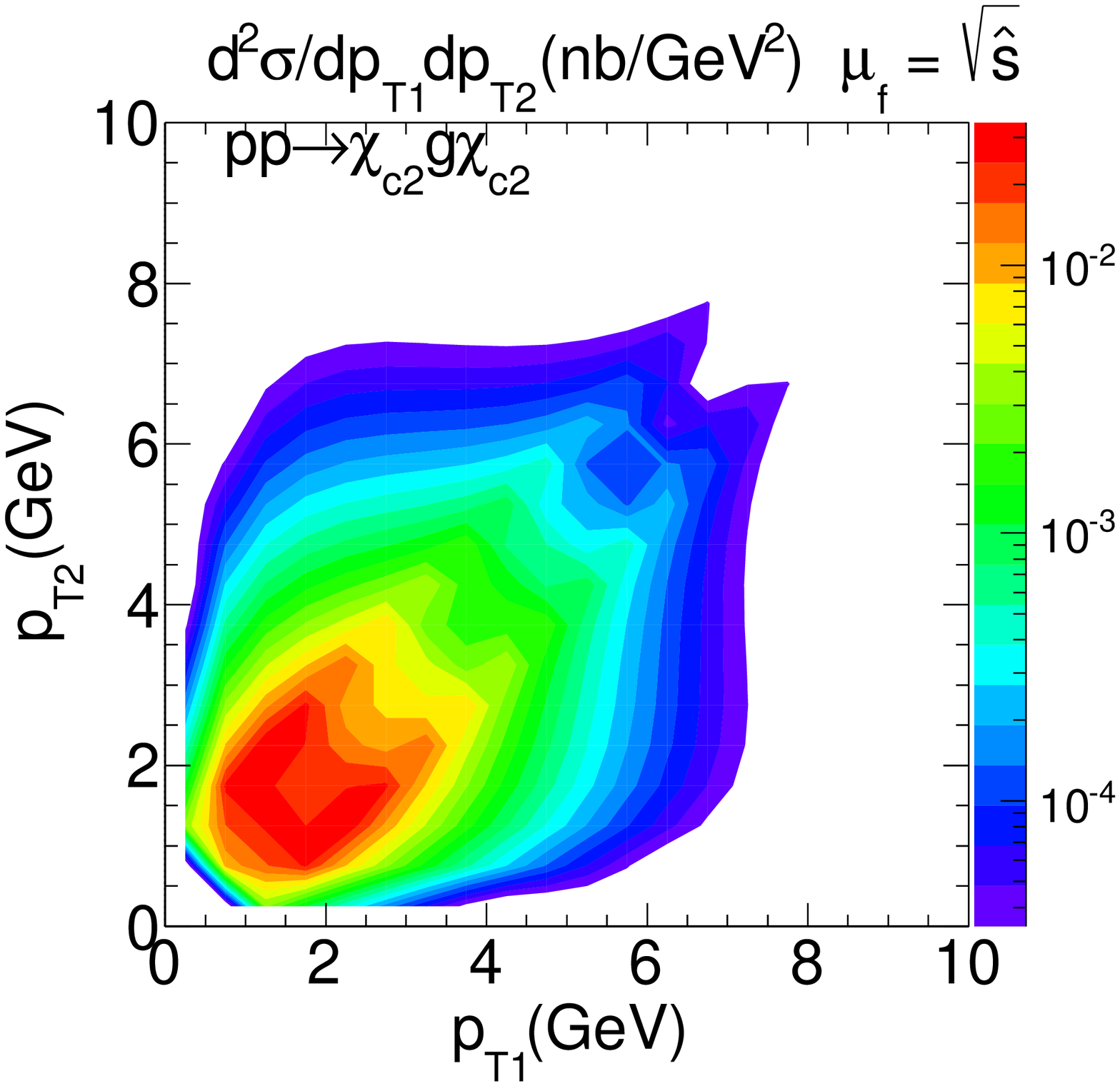}
\caption{Two-dimensional distributions ($p_{T1} \times p_{T2}$)
for $\chi_{c 2} \chi_{c 2}$ production.
Here $\mu_f^2 = \hat s$.
}
\label{fig:p1tp2t_chic2chic2}
\end{figure}

In Fig.\ref{fig:p1tp2t_chic0chic0} -\ref{fig:p1tp2t_chic2chic2} 
we show two-dimensional distributions
in transverse momenta of both $\chi_c$ quarkonia
for separate ((A) or (B) in Fig. \ref{fig:diagrams_chicchicg}) diagrams. 
Such separation is possible due to quite different phase space
population of the different mechanisms (diagrams).

\begin{figure}[h!]
\includegraphics[width= 7.5cm]{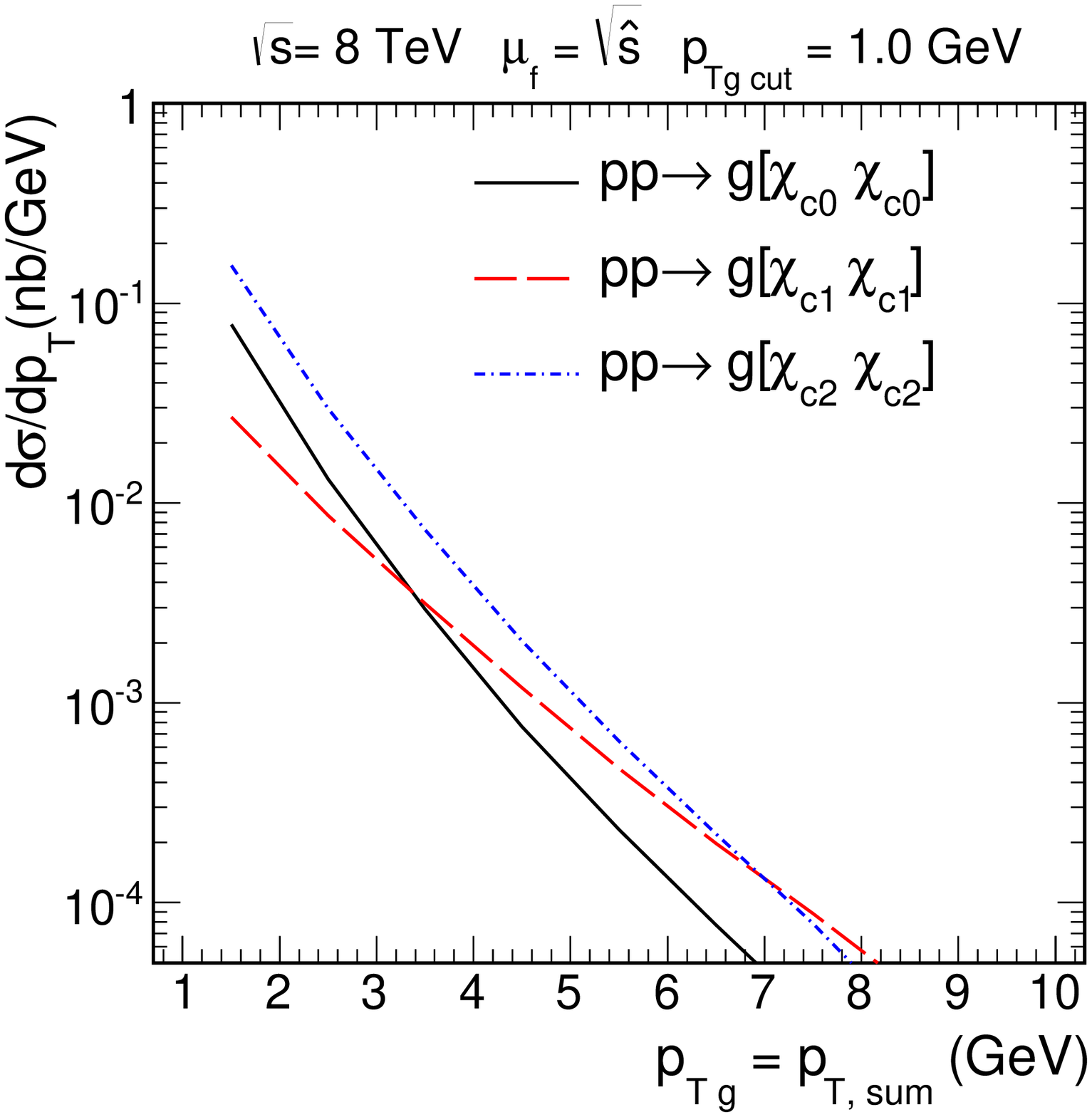}
\includegraphics[width=7.5 cm]{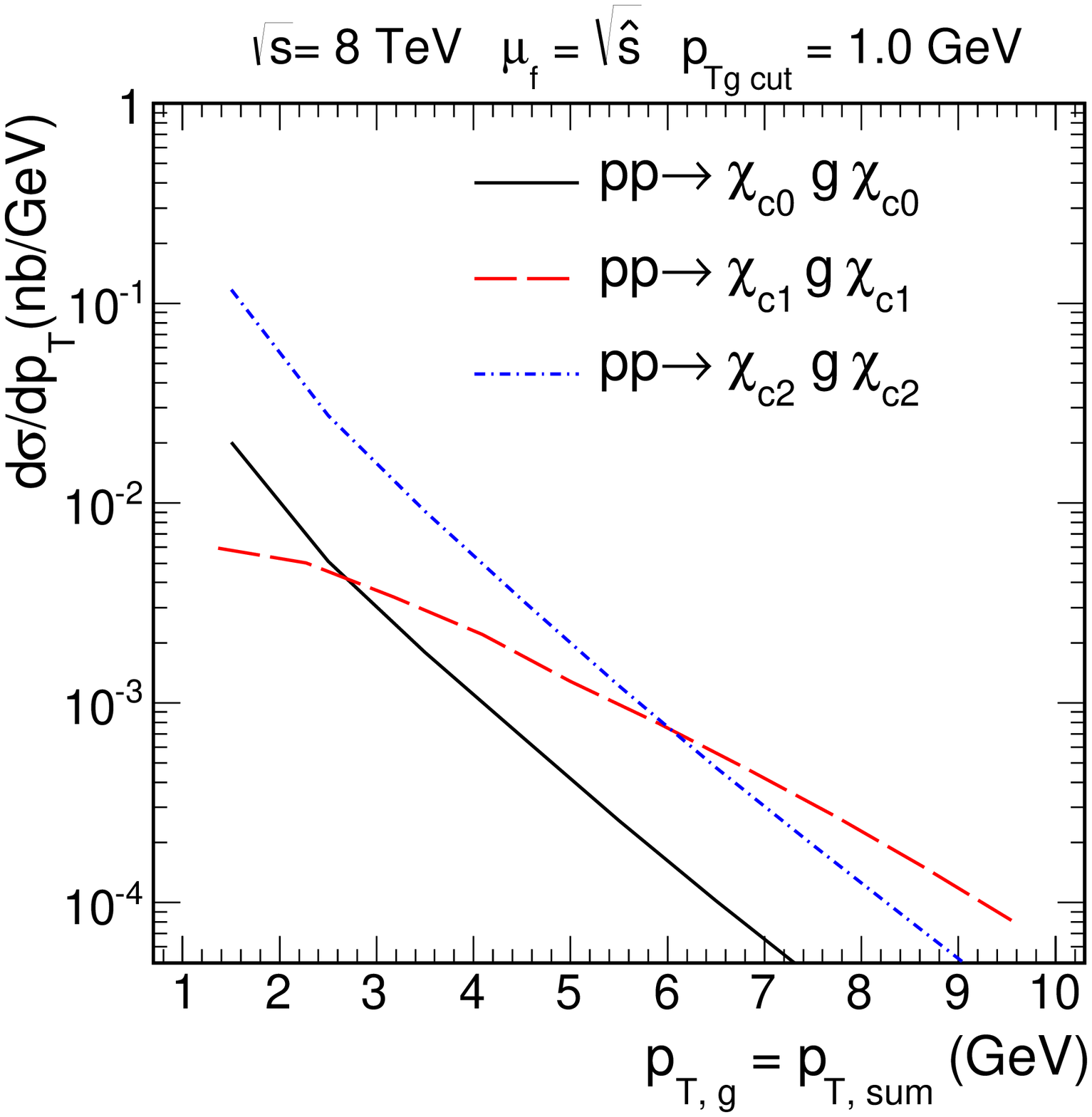}
\caption{Distribution in $\vec{p}_{T,sum}$, where
$\vec{p}_{T,sum}=\vec{p}_{1 T}+\vec{p}_{2 T} $
for $pp\rightarrow g [\chi_{c J} \chi_{c J}]$ (the left panel) and 
$pp\rightarrow \chi_{c J} g\chi_{c J}$ (the right panel) processes for $\sqrt{s}= 8\: \rm{TeV}$.
Here $\mu_f^2 = \hat s$.
}
\label{fig:dsig_dptsum}
\end{figure}

Finally in Fig.\ref{fig:dsig_dptsum} we show distribution in $p_{T,sum}$
(vector sum of transverse momenta of both outgoing quarkonia) for different
involved contributions. Because of the momentum conservation it equals the
transverse-momentum distribution of the emitted gluon.
A significant difference between diagram (A) and (B) or (C) appears for $\chi_{c 1}$.
Emission of the gluon is suppressed  in diagram (A) at small $p_{T}$ region.
While the distributions for $\chi_{c 0} \chi_{c 0}$ and $\chi_{c 2} \chi_{c 2}$
are similar, the distributions for $\chi_{c 1} \chi_{c 1}$ are clearly less
steep. Similar observation was already made in the $k_{T}$-factorization
study in \cite{Cisek:2017ikn}. This is particularly spectacular for the central emission diagram (diagram (C) in 
Fig.\ref{fig:diagrams_chicchicg})  when both gluons are off-mass-shell.

\section{Conclusions}

In the present paper we have calculated differential cross sections
for $\chi_c$ pair production in the collinear approach
including next-to-leading order corrections ($2 \to 3$ processes). 
Here we have considered only symmetric pairs (identical $\chi_c$ mesons).
The present results can be compared to previously calculated
cross sections in the $k_T$-factorization approach with the KMR
unintegrated gluon distributions.
We have found that the leading-order $2 \to 2$ processes
give much smaller cross sections than those in the $k_T$-factorization
approach. Therefore we have calculated higher-order corrections
including $2 \to 3$ processes. There are three typical diagrams
with emission of leading and central gluons (see
Fig.\ref{fig:diagrams_chicchicg}). 
The cross section for leading gluon emission is much larger. 

When adding the leading and (real emission part of the)
next-to-leading order contribution we have obtained results that are similar to the $k_T$-factorization results for the production of $\chi_{c 0} \chi_{c 0}$
and $\chi_{c 2} \chi_{c 2}$ but still considerably less than in the
$k_{T}$-factorization approach for the $\chi_{c 1} \chi_{c 1}$. 
The latter disagreement is likely due to even higher-order (NNLO) contributions 
(involving $2 \to 4$ processes) contained effectively in the $k_T$-factorisation
which may be crucial to include for
the $\chi_{c 1} \chi_{c 1}$ channel as here the vertices
vanish for on-shell gluons. In general, the larger numerical value of 
deviation from the on-shell situation the larger the vertex. We expect that
consistent inclusion of the NNLO corrections may be important
in this particular case and much less important for other cases.
A detailed study will be done elsewhere.

The central gluon emission is interesting in that it enhances production
of $\chi_c$'s at large rapidity distances. This is similar to the Mueller-Navelet production of large rapidity distance dijets and one may think of a larger enhancement from resummation. 

We have calculated several single-particle differential distributions 
in rapidity and transverse momentum of $\chi_c$ mesons as well as
some correlation observables such as two-dimensional distribution
in transverse momenta of both $\chi_c$ quarkonia or in transverse
momentum of the quarkonium pair.

\subsection*{Acknowledgments}
This study was partially supported by the Polish National Science Center grant DEC-
2014/15/B/ST2/02528 and by the Center for Innovation and Transfer of Natural Sciences
and Engineering Knowledge in Rzesz\'ow.



\begin{thebibliography}{100}
\bibitem{review}
  N.~Brambilla {\it et al.},
  Eur.\ Phys.\ J.\ C {\bf 71} (2011) 1534
  [arXiv:1010.5827 [hep-ph]].
\bibitem{GongLiWang:2010}
  B.~Gong, X.~Q.~Li and J.~X.~Wang,
  Phys.\ Lett.\ B {\bf 673}, 197 (2009)
  Erratum: [Phys.\ Lett.\ B {\bf 693}, 612 (2010)]
  [arXiv:0805.4751 [hep-ph]].
\bibitem{CampbellMaltoni:2007}
  J.~M.~Campbell, F.~Maltoni and F.~Tramontano,
  Phys.\ Rev.\ Lett.\  {\bf 98}, 252002 (2007)
  [hep-ph/0703113 [HEP-PH]].

\bibitem{Lansberg:2011}
  J.~P.~Lansberg,
  Phys.\ Lett.\ B {\bf 695}, 149 (2011)
  [arXiv:1003.4319 [hep-ph]].
\bibitem{k_T-fact}
S.~Catani, M.~Ciafaloni and F.~Hautmann,
Nucl.\ Phys.\ B {\bf 366}, 135 (1991);

J.~C.~Collins and R.~K.~Ellis,
Nucl.\ Phys.\ B {\bf 360}, 3 (1991);\newline
E.~M.~Levin, M.~G.~Ryskin, Y.~M.~Shabelski and A.~G.~Shuvaev,
Sov.\ J.\ Nucl.\ Phys.\  {\bf 53}, 657 (1991)
[Yad.\ Fiz.\  {\bf 53}, 1059 (1991)].
\bibitem{Kimber:2001kmr} 
  M.~A.~Kimber, A.~D.~Martin and M.~G.~Ryskin,
  Phys.\ Rev.\ D {\bf 63}, 114027 (2001)
  [hep-ph/0101348].

\bibitem{Baranov:2015yea} 
  S.~P.~Baranov, A.~V.~Lipatov and N.~P.~Zotov,
  Phys.\ Rev.\ D {\bf 93}, no. 9, 094012 (2016)
  [arXiv:1510.02411 [hep-ph]].


\bibitem{Kniehl:2006sk} 
  B.~A.~Kniehl, D.~V.~Vasin and V.~A.~Saleev,
  Phys.\ Rev.\ D {\bf 73}, 074022 (2006)
  [hep-ph/0602179].
\bibitem{Baranov:2002cf} 
  S.~P.~Baranov,
  Phys.\ Rev.\ D {\bf 66}, 114003 (2002).

\bibitem{Baranov:2007dw} 
  S.~P.~Baranov and A.~Szczurek,
  Phys.\ Rev.\ D {\bf 77}, 054016 (2008)
  [arXiv:0710.1792 [hep-ph]].
\bibitem{Cisek:2017gno} 
  A.~Cisek and A.~Szczurek,
  Phys.\ Rev.\ D {\bf 97}, no. 3, 034035 (2018)
  [arXiv:1712.07943 [hep-ph]].
\bibitem{D0_jpsijpsi}
  V.~M.~Abazov {\it et al.} [D0 Collaboration],
  Phys.\ Rev.\ D {\bf 90}, no. 11, 111101 (2014)
  [arXiv:1406.2380 [hep-ex]].
\bibitem{LHCb_jpsijpsi_7TeV}
  R.~Aaij {\it et al.} [LHCb Collaboration],
  Phys.\ Lett.\ B {\bf 707}, 52 (2012)
  [arXiv:1109.0963 [hep-ex]].

\bibitem{Khachatryan:2014iia} 
  V.~Khachatryan {\it et al.} [CMS Collaboration],
  JHEP {\bf 1409}, 094 (2014)
  [arXiv:1406.0484 [hep-ex]].
\bibitem{Aaij:2016bqq} 
  R.~Aaij {\it et al.} [LHCb Collaboration],
  JHEP {\bf 1706}, 047 (2017)
  Erratum: [JHEP {\bf 1710}, 068 (2017)]
  [arXiv:1612.07451 [hep-ex]].
\bibitem{Aaboud:2016fzt} 
  M.~Aaboud {\it et al.} [ATLAS Collaboration],
  Eur.\ Phys.\ J.\ C {\bf 77}, no. 2, 76 (2017)
  [arXiv:1612.02950 [hep-ex]].
\bibitem{Kom:2011bd} 
  C.~H.~Kom, A.~Kulesza and W.~J.~Stirling,
  Phys.\ Rev.\ Lett.\  {\bf 107}, 082002 (2011)
  [arXiv:1105.4186 [hep-ph]].
\bibitem{Luszczak:2011zp} 
  M.~Luszczak, R.~Maciula and A.~Szczurek,
  Phys.\ Rev.\ D {\bf 85}, 094034 (2012)
  [arXiv:1111.3255 [hep-ph]].
\bibitem{sigma_effective}
  F.~Abe {\it et al.} [CDF Collaboration],
  Phys.\ Rev.\ Lett.\  {\bf 79}, 584 (1997)\newline
  F.~Abe {\it et al.} [CDF Collaboration],
  Phys.\ Rev.\ D {\bf 56}, 3811 (1997);\newline
  V.~M.~Abazov {\it et al.} [D0 Collaboration],
  Phys.\ Rev.\ D {\bf 81}, 052012 (2010)
  [arXiv:0912.5104 [hep-ex]];

  G.~Aad {\it et al.} [ATLAS Collaboration],
  New J.\ Phys.\  {\bf 15}, 033038 (2013)
  [arXiv:1301.6872 [hep-ex]];

  S.~Chatrchyan {\it et al.} [CMS Collaboration],
  JHEP {\bf 1403}, 032 (2014)
  [arXiv:1312.5729 [hep-ex]];

  G.~Aad {\it et al.} [ATLAS Collaboration],
  JHEP {\bf 1404}, 172 (2014)
  [arXiv:1401.2831 [hep-ex]];


  R.~Aaij {\it et al.} [LHCb Collaboration],
  JHEP {\bf 1206}, 141 (2012)
  Addendum: [JHEP {\bf 1403}, 108 (2014)]
  [arXiv:1205.0975 [hep-ex]].
\bibitem{Cisek:2017ikn} 
  A.~Cisek, W.~Sch\"afer and A.~Szczurek,
  Phys.\ Rev.\ D {\bf 97}, no. 11, 114018 (2018)
  [arXiv:1711.07366 [hep-ph]].

\bibitem{Baranov:1997ph} 
  S.~P.~Baranov,
  Phys.\ Atom.\ Nucl.\  {\bf 60}, 986 (1997)
  [Yad.\ Fiz.\  {\bf 60}, 1103 (1997)].

\bibitem{Likhoded:2016zmk} 
  A.~K.~Likhoded, A.~V.~Luchinsky and S.~V.~Poslavsky,
  Phys.\ Rev.\ D {\bf 94}, no. 5, 054017 (2016)
  [arXiv:1606.06767 [hep-ph]].

\bibitem{Lipatov:1996ts} 
  L.~N.~Lipatov,
  Phys.\ Rept.\  {\bf 286}, 131 (1997)
  [hep-ph/9610276].
\bibitem{Antonov:2004hh} 
  E.~N.~Antonov, L.~N.~Lipatov, E.~A.~Kuraev and I.~O.~Cherednikov,
  Nucl.\ Phys.\ B {\bf 721}, 111 (2005)
  [hep-ph/0411185].

\bibitem{BFKL} 
V.~S.~Fadin, E.~A.~Kuraev and L.~N.~Lipatov,
Phys.\ Lett.\  {\bf 60B}, 50 (1975);
E.~A.~Kuraev, L.~N.~Lipatov and V.~S.~Fadin,
Sov.\ Phys.\ JETP {\bf 45}, 199 (1977)
[Zh.\ Eksp.\ Teor.\ Fiz.\  {\bf 72}, 377 (1977)];
I.~I.~Balitsky and L.~N.~Lipatov,
Sov.\ J.\ Nucl.\ Phys.\  {\bf 28}, 822 (1978)
[Yad.\ Fiz.\  {\bf 28}, 1597 (1978)].

\bibitem{Mueller:1986ey} 
A.~H.~Mueller and H.~Navelet,
Nucl.\ Phys.\ B {\bf 282}, 727 (1987).

\bibitem{Caporale:2014gpa} 
F.~Caporale, D.~Y.~Ivanov, B.~Murdaca and A.~Papa,
Eur.\ Phys.\ J.\ C {\bf 74}, no. 10, 3084 (2014)
Erratum: [Eur.\ Phys.\ J.\ C {\bf 75}, no. 11, 535 (2015)]
[arXiv:1407.8431 [hep-ph]].

\bibitem{Ducloue:2013bva} 
B.~Duclou\'e, L.~Szymanowski and S.~Wallon,
Phys.\ Rev.\ Lett.\  {\bf 112}, 082003 (2014)
[arXiv:1309.3229 [hep-ph]].
\bibitem{Landau-Yang}
  L.~D.~Landau,
  Dokl.\ Akad.\ Nauk Ser.\ Fiz.\  {\bf 60}, no. 2, 207 (1948);\newline
  C.~N.~Yang,
  Phys.\ Rev.\  {\bf 77}, 242 (1950).


\bibitem{MSTW}
  A.~D.~Martin, W.~J.~Stirling, R.~S.~Thorne and G.~Watt,
  Eur.\ Phys.\ J.\ C {\bf 63}, 189 (2009)
  [arXiv:0901.0002 [hep-ph]].











\end{thebibliography}
\end{document}